\documentclass[10pt]{article}
\usepackage{amsmath,amsthm,latexsym,amssymb,amsfonts,epsfig, psfrag}

\DeclareMathOperator{\diag}{diag}
\DeclareMathOperator{\flow}{\mbox{\boldmath$ \alpha$}}

\makeatletter
\@addtoreset{equation}{section}
\makeatother

\def\be{\begin{equation}}
\def\ee{\end{equation}}
\def\ba{\begin{eqnarray}}
\def\ea{\end{eqnarray}}

\newcommand\nn{\nonumber}
\newcommand\q{\quad}
\def\Nl{{\mathchoice
{\setbox0=\hbox{$\displaystyle\rm N$}\hbox{\hbox to0pt
{\kern0.4\wd0\vrule height0.9\ht0\hss}\box0}}
{\setbox0=\hbox{$\textstyle\rm N$}\hbox{\hbox to0pt
{\kern0.4\wd0\vrule height0.9\ht0\hss}\box0}}
{\setbox0=\hbox{$\scriptstyle\rm N$}\hbox{\hbox to0pt
{\kern0.4\wd0\vrule height0.9\ht0\hss}\box0}}
{\setbox0=\hbox{$\scriptscriptstyle\rm N$}\hbox{\hbox to0pt
{\kern0.4\wd0\vrule height0.9\ht0\hss}\box0}}}}
\def\Zl{{\mathchoice
{\setbox0=\hbox{$\displaystyle\rm Z$}\hbox{\hbox to0pt
{\kern0.4\wd0\vrule height0.9\ht0\hss}\box0}}
{\setbox0=\hbox{$\textstyle\rm Z$}\hbox{\hbox to0pt
{\kern0.4\wd0\vrule height0.9\ht0\hss}\box0}}
{\setbox0=\hbox{$\scriptstyle\rm Z$}\hbox{\hbox to0pt
{\kern0.4\wd0\vrule height0.9\ht0\hss}\box0}}
{\setbox0=\hbox{$\scriptscriptstyle\rm Z$}\hbox{\hbox to0pt
{\kern0.4\wd0\vrule height0.9\ht0\hss}\box0}}}}
\def\Ql{{\mathchoice
{\setbox0=\hbox{$\displaystyle\rm Q$}\hbox{\hbox to0pt
{\kern0.4\wd0\vrule height0.9\ht0\hss}\box0}}
{\setbox0=\hbox{$\textstyle\rm Q$}\hbox{\hbox to0pt
{\kern0.4\wd0\vrule height0.9\ht0\hss}\box0}}
{\setbox0=\hbox{$\scriptstyle\rm Q$}\hbox{\hbox to0pt
{\kern0.4\wd0\vrule height0.9\ht0\hss}\box0}}
{\setbox0=\hbox{$\scriptscriptstyle\rm Q$}\hbox{\hbox to0pt
{\kern0.4\wd0\vrule height0.9\ht0\hss}\box0}}}}
\def\Rl{{\mathchoice
{\setbox0=\hbox{$\displaystyle\rm R$}\hbox{\hbox to0pt
{\kern0.4\wd0\vrule height0.9\ht0\hss}\box0}}
{\setbox0=\hbox{$\textstyle\rm R$}\hbox{\hbox to0pt
{\kern0.4\wd0\vrule height0.9\ht0\hss}\box0}}
{\setbox0=\hbox{$\scriptstyle\rm R$}\hbox{\hbox to0pt
{\kern0.4\wd0\vrule height0.9\ht0\hss}\box0}}
{\setbox0=\hbox{$\scriptscriptstyle\rm R$}\hbox{\hbox to0pt
{\kern0.4\wd0\vrule height0.9\ht0\hss}\box0}}}}
\def\Cl{{\mathchoice
{\setbox0=\hbox{$\displaystyle\rm C$}\hbox{\hbox to0pt
{\kern0.4\wd0\vrule height0.9\ht0\hss}\box0}}
{\setbox0=\hbox{$\textstyle\rm C$}\hbox{\hbox to0pt
{\kern0.4\wd0\vrule height0.9\ht0\hss}\box0}}
{\setbox0=\hbox{$\scriptstyle\rm C$}\hbox{\hbox to0pt
{\kern0.4\wd0\vrule height0.9\ht0\hss}\box0}}
{\setbox0=\hbox{$\scriptscriptstyle\rm C$}\hbox{\hbox to0pt
{\kern0.4\wd0\vrule height0.9\ht0\hss}\box0}}}}
\def\Hl{{\mathchoice
{\setbox0=\hbox{$\displaystyle\rm H$}\hbox{\hbox to0pt
{\kern0.4\wd0\vrule height0.9\ht0\hss}\box0}}
{\setbox0=\hbox{$\textstyle\rm H$}\hbox{\hbox to0pt
{\kern0.4\wd0\vrule height0.9\ht0\hss}\box0}}
{\setbox0=\hbox{$\scriptstyle\rm H$}\hbox{\hbox to0pt
{\kern0.4\wd0\vrule height0.9\ht0\hss}\box0}}
{\setbox0=\hbox{$\scriptscriptstyle\rm H$}\hbox{\hbox to0pt
{\kern0.4\wd0\vrule height0.9\ht0\hss}\box0}}}}
\def\Ol{{\mathchoice
{\setbox0=\hbox{$\displaystyle\rm O$}\hbox{\hbox to0pt
{\kern0.4\wd0\vrule height0.9\ht0\hss}\box0}}
{\setbox0=\hbox{$\textstyle\rm O$}\hbox{\hbox to0pt
{\kern0.4\wd0\vrule height0.9\ht0\hss}\box0}}
{\setbox0=\hbox{$\scriptstyle\rm O$}\hbox{\hbox to0pt
{\kern0.4\wd0\vrule height0.9\ht0\hss}\box0}}
{\setbox0=\hbox{$\scriptscriptstyle\rm O$}\hbox{\hbox to0pt
{\kern0.4\wd0\vrule height0.9\ht0\hss}\box0}}}}
\newcommand{\ca}{\mathcal A}

\newcommand{\ch}{\mathcal H}
\newcommand{\ci}{\mathcal I}

\newcommand{\cn}{\mathcal N}

\newcommand{\cp}{\mathcal P}

\newcommand{\bd}{\mathbf d}

\begin{document}

\title{Gauge invariant perturbations around symmetry reduced sectors of general relativity: applications to cosmology}
\author{Bianca Dittrich\thanks{bdittrich AT perimeterinstitute DOT ca} \\
\it \small Perimeter Institute for Theoretical Physics, 
31 Caroline Street North, Waterloo, ON N2L 2Y5, Canada \\
 Johannes Tambornino\thanks{jtambornino AT perimeterinstitute DOT ca}\\
\it \small Perimeter Institute for Theoretical Physics,
31 Caroline Street North, Waterloo, ON N2L 2Y5, Canada \\
\it \small and\\
\it \small Institut f\"ur Physik, RWTH Aachen, D-52056 Aachen, Germany
}


\maketitle

\begin{abstract}
We develop a gauge invariant canonical perturbation scheme for perturbations around symmetry reduced sectors  in generally covariant theories, such as general relativity. The central objects of investigation are gauge invariant observables which encode the dynamics of the system. 

We apply this scheme to perturbations around a homogeneous and isotropic sector (cosmology) of general relativity. The background variables of this homogeneous and isotropic sector are treated fully dynamically which allows us to approximate the observables to arbitrary high order in a self--consistent and fully gauge invariant manner.  Methods to compute these observables are given.

The question of backreaction effects of inhomogeneities onto a homogeneous and isotropic background can be addressed in this framework. We illustrate the latter by considering homogeneous but anisotropic Bianchi--I cosmologies as perturbations around a homogeneous and isotropic sector.

\end{abstract}


\section{Introduction}\label{intro}

General relativity has two very challenging features: firstly the dynamics of the theory is highly non--linear, secondly general relativity is a diffeomorphism invariant and background independent theory. These two features make it very difficult to construct gauge invariant observables, that is to extract physical predictions.  Diffeomorphism invariance of the theory includes invariance under time reparametrizations, therefore observables have to be constants of motions. Hence finding gauge invariant observables is intimately related to solving the dynamics of the theory. But because of the highly non--linear structure of the theory it is quite hopeless to solve general relativity exactly. Indeed so far there are almost\footnote{For pure gravity only the 10 ADM charges related to the Poincare symmetries are known.} no gauge invariant observables known \cite{torre}. 

Therefore we think that it is important to develop perturbation schemes in order to attack the dynamics of the theory. The difficulty for a perturbation scheme in general relativity is again the control over the gauge dependence of the results , which becomes already quite challinging at second order (see for instance \cite{bruni1}). 

In this work  we will develop a manifestly gauge invariant perturbation theory in the canonical framework. This perturbation theory allows the calculation of observables to an arbitrary high order. In an earlier work we introduced a calculation scheme for gauge invariant observables for perturbations around a fixed background spacetime, which in the canonical framework corresponds to perturbations around a fixed phase space point.  In this work we will rather consider an expansion around a whole symmetry reduced sector of the theory. This sector will be treated non--perturbatively and could for instance describe homogeneous and isotropic fields (i.e. cosmology), spherically symmetric fields (i.e. black holes) or even midi--superspaces such as cylindrically symmetric gravitational waves. The latter example has infinitely many degrees of freedom, but is solvable \cite{kuchar1,torre1}. Note that perturbations around a fixed phase space point arise as a special case of the scheme developed here.

We think that the methods developed here are useful for a quantum theory of gravity as well as for approaching classical dynamics, for instance for cosmological perturbation theory. 

In a quantum theory of gravity gauge invariant (that is physical) observables are the central object. The construction and interpretation of the observables of the theory is therefore one of the key open issues, which is also related to the problem of time, see for instance \cite{kuchar2,rovellibook, bd2,hartle1,bd3,frank}. Approximation methods for gauge invariant observables provide an explicit way to construct observables and may also help to test (quantum) interpretation of these as well as to discuss phenomenological implications \cite{hartle1,bd3}.  The approximation scheme developed in this work shows explictly how to express local observables in a gauge invariant way and therefore indicates how one could understand quantum field theory on curved space time as an approximation to the full theory of quantum gravity. 

Using an approximation scheme for observables around a whole (symmetry reduced) sector of the theory  allows one to explore properties of gauge invariant observables better than in a perturbation scheme around a fixed phase space point. First of all one can now incorporate results from symmetry reduced (exactly solvable) models. The degrees of freedom describing these sectors are treated non--perturbatively. Secondly, as we will see, this approach provides a gauge invariant (perturbative) description of dynamical processes also in closed universes, where for instance the notion of scattering amplitudes is problematic. 

Moreover since the early days of quantum gravity mini-- and midi--superspace models gained by symmetry reduction of the full theory played an important role, especially for describing quantum cosmology (\cite{mini}, \cite{bojo1} and references therein). However because of the non--linear  dynamics of the theory, which leads to a coupling of non--symmetric and symmetric modes,  the question arises whether the results gained from these models are reliable \cite{kuchar3}.   For different attempts to derive symmetry--reduced quantized models from the quantized full theory, see for instance \cite{loll1,cohemb,koslow}. Since the coupling between non--symmetric and symmetric modes arises through the dynamics, we expect that the issues raised here become important if one wants to match solutions to the Hamiltonian constraint in the symmetry reduced model with solutions to the Hamiltonian constraint in the full theory.  We provide such a matching for classical gauge invariant observables, however the observables in the full theory have higher order corrections due to the non--symmetric modes.  These corrections can be given explicitly, thus allowing for an estimation of errors, which arise if one ingnores the non--symmetric modes.

 
Recently there is also growing interest to incorporate (linear) perturbations \cite{MFB} into the framework of loop quantum cosmology \cite{bojo2}. Here we hope that a clear gauge invariant formulation of perturbations of symmetry reduced models might help to understand for instance issues related to the gauge dependence of these approaches. Since the scheme developed in this work is consistent to any order, it proves that linear perturbations can be understood as lowest order of  fully gauge invariant perturbations governed by non--linear dynamics. 
Similar gauge issues are present in recent developments in background independent approaches to the graviton propagator \cite{simone}.

A gauge invariant canonical perturbation theory could be also fruitful in classical applications, such as second (and higher) order perturbation theory around cosmological solutions  \cite{acqua} or black holes. The main difficulty here is to control the gauge dependence of the results. This gauge dependence can be understood from the fact, that one has to identify spacetime points in the ``physical'' (non--symmetric) universe with spacetime points in the ``background'' universe, around which the perturbation is taken. This identification can be related to a choice of coordinates for the ``physical'' universe. One might wonder why we attempt to develop a perturbation theory in the canonical formalism, where one would expect the problem to be even worse due to the foliation for the ``physical'' and ``background'' universe one has to choose in the canonical framework.

The resolution is that we use observables as central objects, i.e. we attempt to approximate directly a gauge invariant observable of the full theory and do not consider (the difference of) fields on two different manifolds representing the perturbed and unperturbed spacetime. Observables in the canonical formalism correspond to phase space functions, gauge invariant observables are invariant under the action of the constraints (the gauge generators).  

The phase space of general relativity is just a representation of the space of all spacetimes (i.e. solutions of the Einstein equations). Gauge invariant phase space functions give the same value on spacetimes which are related by a diffeomorphism. Hence by considering gauge invariant phase space functions we do not need to worry about the identification process between points in the perturbed and unperturbed spacetime.

In order to approximate gauge invariant phase space functions we have to declare which variables are to be considered small. This choice is done in such a way that the approximate observables coincide with the exact observables if evaluated on the symmetry reduced sector of the phase space. Indeed the zeroth order variables can be defined by an averaging procedure. First oder phase space functions vanish on symmetric spacetimes, higher order phase space functions are products of first order phase space functions. Note that the splitting of phase space variables into zeroth and first order is done on the gauge variant level. Generically a gauge invariant phase space function is a sum of terms of different order. 

This approach has similarities to the $(1+3)$ covariant perturbation theory \cite{ellis1} which introduces a preferred timelike observer congruence. In this work we rather have to declare a part of the degrees of freedom, as for instance scalar fields or the longitudinal modes of the gravitational field, to be used as clocks, which might be a more general procedure. 
However we think that it should be possible to recover the $(1+3)$ covariant perturbation theory if one manages to match the choice of clocks to the preferred observer congruence. On the general relation between observables in the canonical theory and ``covariant'' observables, see \cite{bd2}.

A key feature of this work is, that we keep the zeroth order variables as fully dynamical phase space variables and not just as parameters describing the background universe as one would do in a perturbation around a fixed phase space point. Indeed we have to keep the zeroth order variables as canonical variables to allow for a consistent gauge invariant framework to higher than linear order. Moreover this provides a very natural description for backreaction effects: these arise as higher order corrections to observables  arising through averaging of (time evolved) phase space variables.  
Since this approach is gauge invariant it could shed some light on the discussion whether these backreactions are measurable effects or caused by a specific choice of gauge, see for instance \cite{brand,buchert,wald}. As already mentioned we have to choose clocks, which define also the hypersurfaces (by physical criteria, e.g. by demanding that a scalar field is constant on these hypersurfaces) over which the averaging is performed. Therefore the observables describing the backreacion effects depend on the choice of clocks. However, as we will see, one can find relations between the gauge invariant observables corresponding to one choice of clocks and the gauge invariant observables corresponding to another choice of clocks. 

Let us shortly describe the main ideas of the approach developed here: We will approximate a special class of gauge invariant observables, known as complete observables \cite{rovellibook,bd1,bd2}. The complete observables have the advantage that they describe the dynamics of the theory by giving the evolution of certain non--gauge invariant observables (the partial observables) with respect to other non--gauge invariant observables (the clock variables). In \cite{bd1} methods to compute these complete observables were developed, in particular a power series was derived. 

This power series  serves as a starting point for our approximation scheme. We will devide the phase space variables into two sets, namely variables of zeroth and variables of first order. This devision can be implemented by using a projection operator on the space of phase space functions which projects onto the symmetry reduced sector. First order variables are projected to zero. For cosmological applications this projection operator is defined by an averaging procedure.
Now one can define approximate complete observables of order $k$ by omitting in the power series all terms of order higher than $k$. 

Moreover one can define evolution equations for these complete observables and even gauge invariant functions, that generate this evolution (that is Hamiltonians). As we will see the calculation of the complete observables up to a certain order can be cast into a form where ``free propagation'' (i.e. the linear propagation of the first order variables on the symmetry reduced background and the evolution of zeroth order variables in the symmetry reduced sector), is perturbed by interaction processes between first order fields and between first order fields and the zeroth order variables. 

In the next section \ref{appdirac} we will shortly summarize the necessary details concerning complete observables for a general gauge system (in a canonical description). Furthermore we will define gauge invariant Hamiltonians, which generate the physical time evolution with respect to the clocks. Section \ref{appdirac2}  introduces the approximation scheme for a general system around a symmetry reduced sector. There we also define approximate gauge invariant observables of a certain order and consider first properties of approximate complete observables.

In section \ref{cosmo} we consider this scheme for perturbations  around a sector describing isotropic and homogeneous cosmologies. We derive and interpret the power series expressions for the complete observables. 
Then we consider the transformation between complete observables defined with respect to different choices of clocks (which can be understood as representing different families of observers). Section \ref{lapseandshift} defines lapse and shift functions, which allows to compare this canonical approach to covariant approaches. Lapse and shift function determine foliations of spacetimes. These foliations are defined by the choice of clock variables, i.e. by physical conditions.  

Section \ref{clocks} gives explicit examples for clock variables. We consider clocks built from the gravitational degrees of freedom and clocks defined by the matter field degrees of freedom. For the former case we can give a set of clock variables which is related to the longitudinal gauge (see for instance \cite{MFB}), the dynamics of the corresponding complete observables is considered in appendix \ref{longo}. 

In section \ref{scalarmodes}  we consider the equations of motion for first order complete observables (associated to the scalar modes) and a method to find these complete observables. Then we consider the lowest order backreaction effects, that is second order complete observables associated to zeroth order functions. We evaluate the backreaction effect for a very simple toy model, namely a Bianchi--I--universe treated as a perturbation of a homogeneous and isotropic universe, in section \ref{bianchi}. 

The appendix contains a definition of tensor mode decomposition (section \ref{tensor modes}), the equations of motions for the first oder complete observables associated to the scalar modes with longitudinal gauge clocks (section \ref{longo}) and tensor modes (section \ref{gravitons}), as well as a discussion of issues related to the so called linearization instabilities (section \ref{lininstab}).

\section{Complete observables} \label{appdirac}

In this section we will give a short introduction to complete observables for finite dimensional systems. The generalization to infinite dimensional systems is straightforward.  For further details we refer the reader to \cite{bd1,bd2,bd3}. A very short summary of the necessary details on constraint systems can be found in \cite{bd1}, for a longer introduction see for instance \cite{henneaux}. In order to explain some of the properties of complete observables we will express them using a power series, which we are now going to derive.
 
To this end we consider a first class (non--degenerate) constraint system with constraints $\{C_j\}_{j=1}^m$. Since the system is first class  all the constraints generate gauge transformations \cite{henneaux}. Assume that we can find a set of phase space functions $\{T^K\}_{K=1}^m$ (called clock variables) such that the determinant of the matrix
\ba\label{feb031}
A^K_j:=\{T^K\, , \, C_j\}
\ea
is always non--vanishing (as a function on phase space). In this case we can define an equivalent set of first class constraints $\tilde C_K$ by multiplying the original constraints with the inverse of the matrix $A^K_j$:
\ba\label{feb032}  
\tilde C_K:=C_j \,(A^{-1})^j_K  \q 
\ea
where here and in the following we sum over repeated indices.  These new constraints are (weakly, i.e. modulo terms proportional to the constraints) conjugated to the clock variables, that is they satisfy
\ba\label{feb033}
\{T^K,\tilde C_L\} \simeq \delta^K_L
\ea
where by ``$\simeq$'' we denote that this equation holds modulo terms proportional to the constraints. 

From this property (\ref{feb033}) one can prove\footnote
{
Compute $\{\{T^K,\tilde C_M\},\tilde C_L\}$ directly and using the Jacobi identity. Comparing the two results one can conclude that the structure functions $\tilde f_{KM}^L$ defined by $\{\tilde C_K,\tilde C_M\}=\tilde f_{KM}^L\tilde C_L$ have to vanish on the constraint hypersurface.
}
that the constraints $\tilde C_K$ are weakly Abelian, i.e. their Poisson bracket is proportional to terms at least quadratic in the constraints:
\ba\label{feb034}
\{\tilde C_K,\tilde C_L\}=O(C^2) \q .
\ea
Hence the gauge transformations generated by these constraints commute on the constraint hypersurface. This allows us to define gauge invariant observables $F_{[f;T^K]}(\tau)$, called complete observables, by a power series (assuming that this power series converges):
\ba\label{feb035}
F_{[f;T^K]}(\tau)= \sum_{r=0}^\infty \frac{1}{r!} \{\cdots\{f, \tilde C_{K_1}\},\cdots\},\tilde C_{K_r}\}\, (\tau^{K_1}-T^{K_1})\cdots (\tau^{K_r}-T^{K_r})  \q .
\ea
Here $f$ is a phase space function (called the partial observable) and $\{\tau^K\}_{K=1}^m$ are a set of constants. Due to the properties (\ref{feb033},\ref{feb034}) the Poisson bracket of the series (\ref{feb035}) with the constraints vanishes at least weakly. Hence the complete observable $F_{[f;T^K]}(\tau)$ is (weakly) gauge invariant, that is a (weak) Dirac observable.

Furthermore we have that the complete observable coincides with the partial observable $f$ if restricted to the (gauge fixing) hypersurface defined by $\{T^K=\tau^K\}_{K=1}^m$. Therefore we can understand the complete observables as gauge invariant extensions of gauge restricted functions, where the gauge is given by $\{T^K=\tau^K\}_{K=1}^m$. (In this way one can find a definition of complete observables alternative to the power series (\ref{feb035}) , see \cite{bd1}.)
From this observation it follows that associating complete observables to partial observables is an algebraic morphism, i.e. we have
\ba
F_{[f\cdot g+h\, ; T^K]}(\tau) \simeq F_{[f\,;T^K]}(\tau)\cdot F_{[g\,;T^K]}+F_{[h\,T^K]}(\tau)
\ea
as can be also seen in a more elaborate way by using the power series (\ref{feb035}).

The value of the the complete observables is constant on the gauge orbits. In generally covariant systems, such as general relativity, the orbits describing time evolution (i.e. foliations of space--times) are part of the gauge orbits, or in other words coordinate time evolution is actually a gauge transformation. Hence gauge invariant observables are constant in coordinate time.  

Nevertheless one can vary the parameters $\tau^K$ in the complete observable $F_{[f;T]}(\tau)$. Then the value of the phase space function $F_{[f;T]}(\tau)$ {\it does} change. 
 We can interpret this change as a time evolution with respect to the chosen clock variables $T^K$. Indeed the complete observable $F_{[f;T]}(\tau)$ gives the value of the phase space function $f$ at that ``moment'' at which the clocks $T^K$ show the values $\tau^K$. Using the representation of the complete observable as the power series (\ref{feb035}) one can derive the ``equations of motions''
\ba\label{feb036}
\frac{\partial}{\partial \tau^M} F_{[f;T^K]}(\tau) \simeq  F_{[\, \{f,\tilde C_M\}\, ; T^K]}(\tau)
\ea
which replace the usual time evolution equations. The usual initial conditions are now replaced by
\ba\label{feb037}
F_{[f\, ;T^K]}(\tau)_{\; \; {\big{|}}\{T^L=\tau^L\}_{L=1}^m}=f  \q . 
\ea

This closes our short introduction to complete observables. In the remainder of this section we define generalizations of energy like observables.

Here we understand energy as a phase space function which generates the physical time evolution in (one of the parameters) $\tau^M$. Hence we ask whether there exists gauge invariant phase space functions $H_M$ which generate the evolution (\ref{feb036}),  that is these functions should satisfy
\ba\label{cham1}
\frac{\partial}{\partial \tau^M} F_{[f;T^K]}(\tau)  \simeq \{  F_{[f; T^K]}(\tau), H_M(\tau)\}
\ea
where we allowed for a $\tau$--dependence of these $\tau$--generators. If the variation in say $\tau^1$ corresponds to time evolution (with respect to the physical clock $T^1$) one could interpret the corresponding generator $H_1(\tau)$ as a possibly $\tau$--dependent ``physical Hamiltonian'' \cite{tt1} (as opposed to the Hamiltonian or scalar constraint in general relativity which vanishes on physical configurations).

Indeed, in the case that the clock variables $T^K$ Poisson commute, it is possible to find functions $H_K(\tau)$ that satisfy (\ref{cham1}) . However this equation is not satisfied for all phase space functions $f$, we have to exclude functions which depend on the clocks $T^K$ and functions which do not commute with the clocks $T^K$.

We will look for the gauge invariant functions $H_L(\tau)$ by assuming that they can be written as complete observables $F_{[h_L;T^K]}(\tau)$. The partial observables $h_L$ have to satisfy
\ba\label{cham2}
\frac{\partial}{\partial \tau^L} F_{[f;T^K]}(\tau)  \simeq  F_{[\{f,\tilde C_L\}\, ; T^K]}(\tau) \simeq
\{ F_{[f; T^K]}(\tau)\,,\,\,  F_{[h_L;T^K]}(\tau)\} \simeq   F_{[\{f, h_L\}^*\, ; \,\, T^K]}(\tau)
\ea
where $\{f,g\}^*$ denotes the Dirac bracket with respect to the gauge $\{T^K=\tau^K\}$:
\ba\label{cham3}
\{f, h_L\}^* \simeq  \{f,h_L\}-\{f,\tilde C_K\}\{T^K , h_L\}+\{f, T^K\}\{\tilde C_K, h_L\}-\{f,\tilde C_K\}\{T^K,T^M\}\{\tilde C_M, h_L\}   \q .  
\ea

For the last equation in (\ref{cham2}) we use the general property
\ba\label{cham4}
\{ \, F_{[f\, ;\,\, T^K]}\, ,\,\, F_{[g\, ;\,\, T^K]}\, \}\simeq  F_{[\{f, g\}^*\, ; \,\, T^K]}(\tau)
\ea
which can be proven by using the power series (\ref{feb035}). Here one needs to prove equation (\ref{cham4}) only up to terms at least linear in the clocks $T^K$, or in other words on the gauge restricted surface $\{T^K=\tau^K\}$. Then one can use that the right hand side has to be gauge invariant (because the left hand side is), hence it can be written as the gauge invariant extension of a $\{T^K=\tau^K\}$--gauge restricted function. 

Comparing equation (\ref{cham2}) with equation (\ref{cham3}) we see that the functions $h_L$ have to satisfy
\ba\label{cham5}
\{f,\tilde C_L\} &\simeq &\{f,h_L\}-\{f,\tilde C_K\}\{T^K , h_L\}+\{f, T^K\}\{\tilde C_K, h_L\}-\{f,\tilde C_K\}\{T^K,T^M\}\{\tilde C_M, h_L\}  + \nn\\
&& \q  O((T-\tau))  
\ea
where $O(T-\tau)$ denotes terms vanishing on the gauge fixing hypersurface $\{T^K=\tau^K\}$. 

This equation can be satisfied if we chooce $h_L=-P_L$ where $P_L$ is a phase space function such that $\{T^K,P_L\}=\delta^K_L+O(C)$, i.e. $P_L$ has to be a momentum (weakly) conjugated to $T^L$ and to commute with the other clocks. Furthermore we have the condition that $f$ has to commute with $P_L$ as well as with the clocks $\{T^K\}_{K=1}^m$ and that the clocks have to be Abelian. (However the ``momenta'' $\{P_L\}_{L=1}^m$ do not have to Poisson commute with each other.) Note that we can add to $h_L=-P_L$ arbitray functions vanishing on the constraint hypersurface, for instance $\tilde C_L$. Such additions do not change the associated complete observable (on the constraint hypersurface). Also, if one is interested only in the generator associated to one specific $\tau^M$ parameter, say $\tau^1$, it is sufficient to specifiy $P_1$.


Therefore we will assume that the clock variables $\{T^K\}_{K=1}^m$ are Abelian. In this case one can find (locally) symplectic coordinate charts of the form $\{ \{T^K\}_{K=1}^m, \{P_K\}_{K=1}^m,  \{q^a,p_a\}\}$ where the $P_K$ are conjugated to the $T^K$ and the set $\{q^a,p_a\}$ denotes the remaining symplectic coordinates.  

Then we can define complete observables $F_{[-P_L\, ; T^K]}(\tau)$ that generate the evolution in the parameters $\tau^L$ according to 
\ba\label{chami1}
\frac{\partial}{\partial \tau^L} F_{[f\, ; T^K]}(\tau)  \simeq \{ F_{[f\, ; T^K]}(\tau)  , F_{[-P_L\, ; T^K]}(\tau)\} \quad \simeq  F_{[\{f,-P_L\}^*\,;\,\,T^K]}(\tau)
\ea
for $f$ a function of the $\{q^a,p_a\}$ only. (It is actually sufficient that $f$ commutes with the clocks $\{T^K\}_{K=1}^m$ and with the momentum $\Pi_L$.) Note however that the complete observables associated to the set $\{q^a,p_a\}$ give a complete set of gauge invariant observables. Indeed the restriction to these coordinates as partial observables is natural: The complete observable associated to a clock $T^K$ is a constant $\tau^K$, hence we cannot generate evolution in this constant via a Poisson bracket with a generating function.  The canonical momenta $\Pi_K$ can be solved for by using the constraints. 

In general it may happen that the Dirac bracket $\{f, -P_L\}^*$ may depend on the clocks also if $f$ does not depend on the clocks. This will occur if the physical Hamiltonian $F_{[-P_L\, ;\, T^K]}(\tau)$ is $\tau$ dependent. This is analogous to the situation of an explicit time dependent Hamiltonian in usual classical mechanics. As in the latter case we have to add to the time evolution equations a piece that takes care of this $\tau$--dependence (we will also add a piece, which takes care of the possibly dependence of $f$ on the momenta $P_K$):
\ba\label{chami2}
\frac{\partial}{\partial \tau^L} F_{[f\, ; T^K]}(\tau)  \simeq \{ F_{[f\, ; T^K]}(\tau)  , F_{[-P_L\, ; T^K]}(\tau)\} + F_{[ \{f, P_L\} , T^K]}(\tau) + F_{[\{f, T^M\}\{\tilde C_M, P_L\}\, ; T^K]}(\tau)
\ea  
holds for arbitrary phase space functions $f$.

A $\tau$--generator $F_{[-P_L\, ;\, T^K]}(\tau)$ does not depend on $\tau$ if the momentum $P_L$ Poisson commutes with the constraints, i.e. if $P_L$ is gauge invariant. This corresponds to a time independent Hamiltonian in classical mechanics, i.e. systems with conserved energy. One can always find a set of Abelian clocks $\{T^K\}_{K=1}^m$ and a set of conjugated momenta $\{P_K\}_{K=1}^m$ such that $l=\text{min}(m,n-m)$ of the momenta are gauge invariant and do not vanish identically on the constraint hypersurface.\footnote
{
This statement has to be understood locally, i.e. the phase space function in question may only  be defined locally. The proof  \cite{phD} basically uses that one can locally always find a polarization of the phase space in which $(n-m)$ of the symplectic coordinate pairs are gauge invariants \cite{gaugebook}. }
  Here $2n$ is the dimension of the phase space. 

We want to emphasize that given a set of clocks the $\tau$--generators $F_{[-P_L\, ;\, T^K]}(\tau)$ are {\it not} uniquely determined. To have uniquely determined $\tau$--generators (modulo constants and constraints) we have to specify the momenta conjugated to the clocks  or alternatively the partial observables we want to evolve according to (\ref{chami1}). One choice of momenta would be $P_L=\tilde C_L$. In this case the $\tau$--generators would vanish on the constraint hypersurface. This is consistent with equation (\ref{chami1}) because functions $f$ allowed in this equations have to be gauge invariant, which means that the associated complete observables do not depend on the parameters $\tau^K$. 

In summary we see that complete observables can be used to express dynamics in a gauge invariant way and lead furthermore to a generalization of the notions of time (which is specified by using clocks) and energy  (which is specified by the clocks and a subset of partial observables for which we want to describe dynamics).

\section{Approximate complete observables}\label{appdirac2}

We will now develop a perturbation scheme for the complete observables. For such a scheme we need to specify which kind of quantities we perturb in, i.e. the quantities assumed to be small. In this approach these quantities will be phase space dependent, therefore we will get a good approximation in certain regions of phase space. One example of such quantities are deviations from a fixed phase space point (serving as a background, e.g. a phase space point describing flat space). We considered a perturbation scheme for this case in \cite{bd3}. 

In this work we are interested in deviations from a whole (symmetry reduced) sector of the phase space. In general relativity such a sector could for instance correspond to homogeneous and isotropic space--times. In general we will describe the sector, we want to perturb around, by a linear projection operator $\cp$, which acts on the space of phase space functions. The projection property means that $\cp \cdot \cp =\cp$.

In particular we can apply the projection operator to some set of symplectic coordinates (which can be understood as phase space functions) $(\chi^a ,\pi_a)$ with $a=1,\ldots,n$ where $2n$ is the dimension of the phase space. Then we can write 
\ba\label{feb038}      
\chi^a=\cp \cdot \chi^a  \,+\, (\text{Id}-\cp)\cdot \chi^a  \q,\q\q\q\q  \pi^a=\cp \cdot \pi^a \, +\, (\text{Id}-\cp)\cdot \pi^a   \q .
\ea
The sector we perturb around is given by the vanishing of the ``fluctuations'' $(\text{Id}-\cp)\cdot \chi^a$ and $(\text{Id}-\cp)\cdot \pi^a$. 

For our purposes we will assume that the projection operator $\cp$ is of the following form\footnote{This assumption can be cast into the language of Poisson embeddings, see \cite{koslow}. It ensures that the kinematics of the symmetry reduced system and of the symmetry reduced sector embedded into the full phase space coincide.}:  There exist symplectic coordinate charts $({\chi'}^c,{\pi'}_c)_{c=1}^n$ such that the projection on the symplectic coordinates is given by
\ba\label{feba}
\cp\cdot {\chi'}^a=0 \q,\q\q\q  \cp\cdot {\pi'}^a=0
\ea
for $a$ in some subset $I$ of the index set $\{1,\ldots,n\}$ and 
\ba\label{febb}
\cp\cdot {\chi'}^b={\chi'}^b \q,\q\q\q  \cp\cdot {\pi'}^b={\pi'}^b
\ea
for the remaining indices $b \notin I$.

The projection operator acts on a general phase space function $f$ by setting all fluctuation variables $({\chi'}^a,{\pi'}_a)_{a\in I}$ in $f$ to zero, that is 
\ba
(\cp\cdot f)({\chi'}^a,\,{\pi'}_a ;\, {\chi'}^b,\,{\pi'}_b  )=f(\cp\cdot {\chi'}^a,\,\cp\cdot {\pi'}_a ;\, \cp\cdot {\chi'}^b,\,\cp\cdot {\pi'}_b    )=  f(0,\,0 ; {\chi'}^b,\,{\pi'}_b      ) \q .
\ea
where the index $a$ takes values in the set $I$ and the index $b$ in $\{1,\ldots,n\}/I$ 

The  fluctuation variables $({\chi'}^a,{\pi'}_a)_{a\in I}$ will be considered to be small, that is functions linear in these variables are defined to be of first order, functions quadratic in these variables of second order and so on. The variables $({\chi'}^b,{\pi'}_b)_{b\notin I}  $ are defined to be of zeroth order. 

It is sometimes more convenient not to work with the symplectic coordinates $({\chi'}^c,\,{\pi'}_c)_{c=1}^n$ but with some other set $(\chi^c,\pi_c)$. The projection operator acts of course also on this set and we can define the (degenerate) coordinates
\ba\label{feb041}
x^c=(\text{Id}-\cp)\cdot \chi^c \q,\q\q p_c=(\text{Id}-\cp)\cdot \pi_c  \q,\q\q X^c=\cp\cdot \chi^c \q,\q\q P^c=\cp\cdot \pi_c  \q .
\ea
Since $\chi^c=X^c+x^c$ and $\pi_c=P_c+p_c$ we can expand (suitable) phase space functions in the fluctuation variables $(x^a,p_a)$ and introduce a classification of phase space functions by defining the variables $(x^a,p_a)$ to be of first order. This classification will coincide with the previous one if the coordinates $({\chi'}^c,\,{\pi'}_c)_{c=1}^n$ and $(\chi^c,\pi_c)$ are related by a linear symplectic transformation, which we will assume to be the case.

 The coordinates (\ref{feb041}) do not need to be symplectic anymore, the Poisson brackets have to be determined by 
\ba\label{feb039} 
\{X^c,P_d\}=\{ \cp \cdot\chi^c\,,\, \cp\cdot \pi_d \}
\q,\q\q 
\{x^c,p_d\}=\{ (\text{Id}-\cp)\cdot \chi^c\,,\, (\text{Id}-\cp)\cdot \pi_d \} 
\q .
\ea
The Poisson brackets between fluctuation variables $(x^c,p_c)$ and the ``sector'' variables $(X^c,P_c)$ vanish. Moreover we have the condition on the fluctuation variables that $\cp \cdot x^a=\cp\cdot p_a=0$. Also the coordinates $(X^a,P_a)$ will be in general highly degenerate.

The Poisson bracket of a phase space function of order $l$ with a phase space function of order $k$ can in general consist of a term of order $(l+k)$ and of a term of order $(l+k-2)$ (for $l,k \geq 1$). The higher order term can arise through the Poisson bracket between the zero order variables $(X^c,P_c)$. The Poisson bracket of a phase space function of order $l$ with a zeroth order term is of order $l$ (or vanishes).

Note that perturbations around a fixed phase space point $m_0$ arise as a special case: The projection operator is given by $\cp\cdot f = f(m_0)\cdot 1$, that is $\cp$ maps all functions to constant functions, where the constant is given by the evaluation of the phase space function at the phase space point $m_0$. Symplectic coordinates $({\chi'}^c, {\pi'}_c)$ with properties (\ref{feba},\ref{febb}) can be found starting from any symplectic coordinate chart $(\chi^c,\pi^c)$ and defining ${\chi'}^c=\chi^c-\chi^c(m_0)$ and $\pi'_c=\pi_c-\pi_c(m_0)$. The index set $I$ in (\ref{feba},\ref{febb}) coincides with the set $\{1,\ldots,n\}$.

For the following we will introduce some notation in order to specify terms of a certain order in a phase space function $f$: with ${}^{(k)}f$ we will denote all terms which are of order $k$ in $f$, with ${}^{[k]}f$ we will denote all terms in $f$ which are of order less or equal to $k$.

We define gauge invariant observables of order $k$ as phase space function which commute with the constraints modulo terms of order $k$ (and modulo constraints). Gauge invariant phase space functions of order $k$ can be obtained from phase space functions $F$ which are exactly gauge invariant by omitting all terms of order higher than $k$, i.e. by truncating to ${}^{[k]}F$:
\ba
\{{}^{[k]}F, C_j\}=\{F,C_j\}\,+\, \{O(k+1)\,,\, {}^{(0)}C_j+{}^{(1)}C_j+ \cdots \} \simeq O(k)  \q .
\ea
where by $O(l)$ we denote terms of order $l$ or higher. Here the lowest order term on the right hand side will in general appear through the Poisson bracket of the $O(k+1)$ term with the first oder term ${}^{(1)}C_j$ of the constraint. All other terms are of higher order.   

In particular we can find approximate complete observables of order $k$ by considering their truncation to order $k$. In the following  we will assume that the constraints $\tilde C_K$ and the clocks $T^K$ can be divided into two subsets $\{\{\tilde C_H\}_{H \in \ch}, \{\tilde C_I\}_{I \in \ci}\}$ and $\{\{T^H\}_{H \in \ch}, \{T^I\}_{I \in \ci}\}$, such that the clocks $T^H$ are of zeroth order and the clocks $T^I$ are of first order. For the constraints  $C_H$ we assume that for an arbitrary first order function ${}^{(1)}f$
\ba\label{fassump}
\{{}^{(1)}f, \tilde C_H\}= O(1)   \q  ,
\ea
which is satisfied if the constraints $\tilde C_H$  do not have a first order term ${}^{(1)}\tilde C_H=0$, however they may have a zeroth order term. For the constraints $\tilde C_I$ we will assume that the zeroth order terms vanish and that the first order terms do not vanish. Consider the power series for the complete observables (\ref{feb035})
\ba\label{feb045}
F_{[f;T^K]}(\tau)
 &\simeq& 
\sum_{r=0}^\infty \frac{1}{r!} \{\cdots\{f, \tilde C_{K_1}\},\cdots,\},\tilde C_{K_r}\}\, (\tau^{K_1}-T^{K_1})\cdots (\tau^{K_r}-T^{K_r}) 
\nn\\
  &\simeq&
\sum_{r=0}^\infty \sum_{s=0}^r \frac{1}{(r - s)!s!} \{\cdots\{f,\tilde C_{H_1}\},\cdots\},\tilde C_{H_{(r-s)}}\},\tilde C_{I_1}\},\cdots\},\tilde C_{I_s}\}\times   \nn\\
&& \q\q\q\q\q\q
(\tau^{H_1}-T^{H_1})\cdots (\tau^{H_{(r-s)}}-T^{H_{(r-s)}}) \times (\tau^{I_1}-T^{I_1})\cdots (\tau^{I_{s}}-T^{I_{s}})
 \nn\\
&\simeq&
\sum_{p=0}^\infty \sum_{q=0}^\infty \frac{1}{p!\, q!}
\{\cdots\{f,\tilde C_{H_1}\},\cdots\},\tilde C_{H_q}\} (\tau^{H_1}-T^{H_1})\cdots (\tau^{H_{q}}-T^{H_{q}})\, \,,
 \tilde C_{I_1}\},\cdots\},\tilde C_{I_p}\}\times 
\nn\\
&& \q\q\q\q\q\q
(\tau^{I_1}-T^{I_1})\cdots   (\tau^{I_{p}}-T^{I_{p}})
\ea
where we used that we can rearrange the constraints in any order and that the clocks $T^H$ commute (at least weakly) with the constraints $\tilde C_I$. Now set the parameters $\tau^I$ to zero. Then for the zeroth order complete observable associated to a zeroth order function ${}^{(0)}f$ we have
\ba\label{feb046}
{}^{(0)} F_{[f;T^K]}(\tau^H,\tau^I=0)&\simeq&
 \sum_{q=0}^\infty \frac{1}{ q!}
{}^{(0)}(\{\cdots\{{}^{(0)}f,\tilde C_{H_1}\},\cdots\},\tilde C_{H_q}\}) (\tau^{H_1}-T^{H_1})\cdots (\tau^{H_{q}}-T^{H_{q}})\nn\\ 
&\simeq&
 \sum_{q=0}^\infty \frac{1}{ q!}
\{\cdots\{{}^{(0)}f,   {}^{(0)}   \tilde C_{H_1}\},\cdots\},{}^{(0)}   \tilde C_{H_q}\} (\tau^{H_1}-T^{H_1})\cdots (\tau^{H_{q}}-T^{H_{q}}) \q   \nn\\
\ea
where the second equation holds due to our assumption on the constraints $\tilde C_H$ to have vanishing first order parts. There only appear zeroth order variables in the second line in (\ref{feb046}), hence we can say that the zeroth order complete observables associated to a zeroth order function\footnote{
Note that the complete observables associated to a zeroth order function have vanishing first order, hence the zeroth complete observable coincides with the first order complete observable in this case. The zeroth order complete observable associated to a zeroth order function is therefore a gauge invariant observable of first order.
}
 are the complete observables of the symmetry reduced sector. The next higher order correction to this complete observable is a second order term and can be considered as the correction (backreaction) term to the dynamics of the symmetry reduced sector due to deviations from symmetry (in the initial values). 

One can also consider for instance the first order complete observable associated to a first order function. As we will see these observables describe the propagation of linear perturbations (which are linearly gauge invariant) on the symmetry reduced sector.   

Note that this approach allows to find gauge invariant observables to any order $k$ by omitting in the series for the complete observables all terms of order higher than $k$. For this the assumptions we made on the clocks $\{\{T^H\}_{H \in \ch}, \{T^I\}_{I \in \ci}\}$ and the constraints $\{\{\tilde C_H\}_{H \in \ch}, \{\tilde C_I\}_{I \in \ci}\}$ are not strictly necessary. However we will see that with these conditions the computation of the complete observables is similar to the usual perturbative calculations involving the ``free'' propagation of perturbations and their interaction as well as the interaction of the zeroth order variables with the perturbations.   

If the power series (\ref{feb045}) for the complete observable converges it defines an exact gauge invariant observable which coincides with the approximate Dirac observable ${}^{[k]}F_{[f;T^K]}(\tau)$ modulo terms of order $(k+1)$. If the power series does not converge in some phase space region, this might be due to the fact that the clock variables $T^K$ do not provide a good parametrization of the gauge orbits in this phase space region \cite{bd1}. 
In this case one can try to find a set of new clock variables ${T'}^K$,  with a better behaviour in this respect and such that (\ref{fassump}) is satisfied also for these new clocks. Assume that the complete observable  $F_{[f';\,{T'}^K]}({\tau'}^H,{\tau'}^I=0)$ associated to these new clock variables and the partial observable $f':={}^{[k]}F_{[f;T^K]}(\tau^H,\tau^I=0)$ can be defined. This complete observable will also coincide with ${}^{[k]}F_{[f;T]}(\tau )$ modulo terms of order $k$, 
as can be seen by examining the power series (\ref{feb045}) for a complete observable and using that $f'$ Poisson commutes modulo terms of order $k$ with the constraints.

\section{Application to cosmology} \label{cosmo}

We will now apply the formalism to general relativity. The sector we perturb around will be the sector describing homogeneous and isotropic (FLRW) cosmologies with a minimally coupled scalar field. We will work with (complex) connection variables \cite{ash,MQG}, however the formalism is independent of the choice of variables and can be also applied to real connection variables or the metric (ADM) variables \cite{ADM}.  We will follow the conventions in \cite{MQG}.

The canonical variables are fields on a spatial manifold $\Sigma$ the coordinates of which we will denote by $\{\sigma^a\}_{a=1}^3$. We will assume that $\Sigma$ is a compact manifold and diffeomorphic to the 3--Torus $T^3=S^1 \times S^1\times S^1$ (in other words the fields are assumed to be periodic). For convenience we will assume that the coordinate length of each spatial direction is equal to $L=1$.  

 The configuration variables are given by a (complex) connection $\{A^j_a\}_{j,a=1}^3$ where latin letters from the beginning of the alphabet denote spatial indices and from the middle of the alphabet $su(2)$--algebra indices:
\ba\label{E1}
A^j_a=\Gamma^j_a+\beta K^j_a
 \q .
\ea
We denote by $\beta=i/2$ the Immirzi parameter, $\Gamma^j_a$ is the spin connection for the triads $e^j_a$ and $K^j_a=2K_{ab}e^b_j$ is the extrinsic curvature where $e^b_j$ is the inverse to the triad $e^j_b$. The spatial metric can be calculated from the triads by $q_{ab}=e^j_a e^k_b\delta_{jk}$. The conjugated momenta  $E^a_j$ are constructed out of the triads
\ba\label{E2}
E^a_j=\beta^{-1} \epsilon^{a a_1 a_2}\epsilon_{jj_1j_2}e^{j_1}_{a_1}e^{j_2}_{a_2}
\ea
where $\epsilon^{a a_1 a_2}$ and $\epsilon_{jj_1j_2}$ are totally anti--symmetric tensors with $\epsilon_{123}=\epsilon^{123}=1$. This gives the following relation between the momenta $E^a_j$ and the 3--metric
\ba
\text{det}(q_{cd})\, q^{ab}=\beta^2 E^a_jE^b_k \delta^{kj} \q .
\ea

The Poisson brackets between the phase space variables are
\ba\label{E3}
\{A^j_a(\sigma),E_k^b(\sigma')\}=\kappa \delta^j_k \delta_a^b \delta(\sigma,\sigma')
\ea
where $\kappa=8\pi G_N/c^3$ is the gravitational coupling constant. 

Furthermore we have a scalar field $\varphi$ and its conjugated momentum $\pi$ which satisfy the commutation relation
\ba
\{\varphi(\sigma),\pi(\sigma')\}=\gamma \delta(\sigma,\sigma')
\ea
where $\gamma$ is the coupling constant for the scalar field.

The constraints are given by the Gauss constraints $G_j(\sigma)$, the vector constraints $V_a(\sigma)$ and the scalar constraints $C(\sigma)$:
\ba\label{E4}
G_j &=& \kappa^{-1}(\partial_a E^a_j+\epsilon_{jkl}A^k_aE^a_l) \nn\\
V_a  &=&  \kappa^{-1}F^j_{ab}E^b_j + \gamma^{-1}\pi \partial_a \varphi
=\kappa^{-1}(\partial_a A_b^j-\partial_bA^j_a +\epsilon_{jkl}A_a^kA_b^l)E^b_j
+ \gamma^{-1}\pi \partial_a \varphi
\nn\\
C &=&  \kappa^{-1} 
\beta^2 F^j_{ab}\epsilon_{jkl} E^a_k E^b_l 
+\gamma^{-1}(\tfrac{1}{2}\pi^2+\tfrac{1}{2}\beta^2 E^a_jE^b_j \partial_a \varphi \partial_b\varphi + \beta^3 \text{det}(E^a_j)V(\varphi) )  \q 
\ea
where 
$V(\cdot)$ is the potential for the scalar field.
Note that we use the scalar constraint with density weight two here.
In the following it will be convenient to work with the following combination of the constraints: 
\ba
G_b &:=&\beta\delta^j_b G_j\\
D_a &:=& V_a-A_a^j G_j \\
S &:=& C+2 \beta^2\,\partial_a(E^a_j G_k \delta^{jk})  \q .
\ea
In particular $D_a$ is now quadratic in the canonical variables and acts as a diffeomorphism constraint.

We will expand the canonical variables around homogeneous and isotropic field configurations in the following way:
\begin{xalignat}{2}\label{cosmo1}
&{A_a}^j(\sigma) =A\,\beta \delta_a^j+{a_a}^b(\sigma)\,\beta\delta^j_b \q ,\q\q 
&&{E^a}_j(\sigma) = E\,\beta^{-1}\delta^a_j +{e^a}_b(\sigma)\,\beta^{-1} \delta^b_j
  \nn\\
&\varphi(\sigma) \q =\Phi+\phi(\sigma) \q, \q\q  
&&\pi(\sigma)\q =\Pi +\rho(\sigma)       \q .
\end{xalignat}
In this way $A,E$ are real if evaluated on a homogeneous cosmology with flat slicing. This division of the phase space into a homogeneous--isotropic sector and an inhomogeneous sector can be implemented by a projection operator $\cp$ acting on the fields defined by:
\begin{xalignat}{2}\label{proj}
& A\,\beta  =
\cp \cdot {A_a}^j 
:=\frac{1}
{3
}\int_\Sigma \delta^a_j\,{A_a}^j \,\bd\sigma  
&&
E\,\beta^{-1} 
=  \cp\cdot {E^a}_j 
:= \frac{1}
{3
}\int_\Sigma \delta^j_a\,{E^a}_j \,\bd\sigma 
\nn \\
& \Phi \q = \cp\cdot\varphi \q :=
\int_\Sigma\varphi\, \bd\sigma
&& \Pi  \q\q = \cp\cdot \pi \q := 
\int_\Sigma\pi\, \bd\sigma   
\end{xalignat}
The Poisson brackets between the homogeneous variables and between the fluctuation variables can be found by using the projection (\ref{proj}):
\begin{xalignat}{2}\label{poisfeb}
&
\{A,E\}=\frac{\kappa}{3
}    
&&   
\{{a_a}^b(\sigma),{e^c}_d(\sigma')\}=\kappa\delta_a^c\delta^b_d\delta(\sigma,\sigma')-\frac{\kappa}
{3
} \delta_a^b\delta^c_d  \nn\\
& 
\{\Phi,\Pi\} =\gamma
&& 
\{\phi(\sigma),\rho(\sigma')\} \q\q\!\! =\gamma\delta(\sigma,\sigma')
-
\gamma  \q\q .
\end{xalignat}
The Poisson bracket between a homogeneous variable and a fluctuation variable vanishes.

In the following we will raise and lower the indices with the Kronecker symbols $\delta^{ab}$ or $\delta_{ab}$ respectively (and not with the background metric $Q_{ab}:=E\delta_{ab}$).

It will be convenient to work with the Fourier transformed variables, using these the Poisson bracket relations simplify. For any field  $f(\sigma)$ we define
\ba
f(k)=
\int_\Sigma  \exp(-i k\cdot\sigma) f(\sigma) \bd\sigma
\ea
where $k\cdot\sigma:=k_a\sigma^a$ and  the wave vector $k$ takes values in $
2\pi
\,\Zl^3$. The inverse transform is
\ba
f(\sigma)= \sum_{k \in \{
2\pi
\,\Zl^3 \} } \exp(i k\cdot\sigma) f(k)  \q .
\ea
Now the homogeneous variables are given by the ($\frac{1}{3}\times$ trace of the) $k=0$ modes of the fields. The Poisson brackets for the Fourier modes of the fluctuation variables are
\ba
\{a_{ab}(k),e^{cd}(k')\}& =&
\kappa
\delta_a^c \delta_b^d\delta_{k,-k'} -
\frac{\kappa}{3}
\delta_{ab}\delta^{cd}\delta_{k,0}\delta_{k',0} \nn\\
\{\phi(k),\rho(k')\} &=& 
\gamma
\delta_{k,-k'} - 
\gamma
\delta_{k,0}\delta_{k',0} \q   \q\q\q\q
\ea
where the additional terms on the right hand side implement that ${a_a}^a(0)={e^a}_a(0)=\phi(0)=\rho(0)=0$.

We want to remark that the Fourier transformed variables can be used to define the symplectic coordinates used in section \ref{appdirac2} in which the projection operator $\cp$ maps part of the symplectic coordinates to zero and leaves the other coordinates invariant. The homogeneous part of the coordinates would be given by $(\sqrt{3}A,\sqrt{3}E;\Phi,\Pi)$. The symplectic pairs that are mapped to zero are given by $(a_{ab}(k),e^{ab}(-k))$ and $(\phi(k),\pi(-k))$ for $k\neq 0$ and 
\ba
&&\left(\sqrt{\tfrac{2}{{3}}}(\tfrac{1}{2}a_{11}(0)-a_{22}(0)+\tfrac{1}{2}a_{33}(0))\,\, ,\,\,
 \sqrt{\tfrac{2}{{3}}}  (\tfrac{1}{2}e^{11}(0)-e^{22}(0)+\tfrac{1}{2}e^{33}(0))\,   \right) \nn\\ 
&&\left({\tfrac{1}{{\sqrt{2}}}}(a_{11}(0)-a_{33}(0))\,\, , \,\, \tfrac{1}{\sqrt{2}}(e^{11}(0)-e^{33}(0))\right) \q .
\ea

We expand the constraints $C_j$ (where $C_0=S,\, C_{Da}=D_a,\, C_{Ga}=G_a$) in the homogeneous and fluctuation variables in order to find the $m$-th order parts ${}^{(m)}C_j$, taking the fluctuation variables as first order and the homogeneous variables as zero order quantities.
The zeroth order of the constraints vanishes except for the scalar constraint:
\ba\label{cosmo3}
{}^{(0)}S &=&\kappa^{-1}
6 \beta^2 A^2 E^2 
+
\gamma^{-1}
\big(
\frac{1}{2} \Pi^2 +E^3 V(\Phi)
\big)  \q .
\ea
The first order parts of the constraints are given by
\ba\label{cosmo4}
{}^{(1)}G_b &=& \kappa^{-1}
\big( 
\partial_a{e^a}_b+\beta\, A\,{\epsilon_{bac}}  {e^{ac}} + \beta\, E\,{\epsilon_{bca}}  {a}^{ac} 
\big) 
\nn \\
{}^{(1)}D_a &=& \kappa^{-1}
\big(
E\,(\partial_a {a_b}^b -\partial_b {a_a}^b) -A \,\partial_b {e^b}_a
\big) +
 \gamma^{-1}\, \Pi\, \partial_a \phi  \nn \\
{}^{(1)}S &=& \kappa^{-1}  
\big(
2\,E\, \partial_b\partial_a e^{ab} + 2 \beta \,A E\,\epsilon_{bac}\partial^b e^{ac}  + 4\beta^2\, A E^2 \,{a_b}^b +
4 \beta^2\, A^2 E\,  
{e^a}_a  
\big)+ \nn\\
&&
 \gamma^{-1}\big(  \Pi\, \rho + E^3 V'(\Phi)\, \phi +E^2 V(\Phi)\, {e^a}_a \big)
\ea
where in the Fourier transformed quantities the partial derivative stands for $(\partial_a f)(k)=ik_a f(k)$. Note that the first order of the zero modes of the diffeomorphism constraints $D_a$ vanish, this is related to the linearization instabilities of backgrounds with compact spatial slices and Killing vectors \cite{fischer}. In the following we will ignore the integrated diffeomorphism constraints $D_a(0)$ and show in appendix \ref{lininstab} that one can indeed deal with these constraints after one has computed the complete observable with respect to all the other constraints. Another way to circumvent the problem of linearization instabilities is to couple the system to massless scalar fields and perturb around non--homogeneous field configurations of these scalar fields. This will be explained in section \ref{scalar}.
 
 The first order part of the zero mode scalar constraint vanishes too, however the overall scalar constraints has a zero order component. For this reason we do not have a linearization instability corresponding to the scalar constraint: rather than viewing the second order part of the integrated scalar constraint as a restriction on the first order variables, we see it as a correction to the zeroth order part, signifying a backreaction effect of the perturbation variables onto the homogeneous variables. This viewpoint is possible because the zeroth order variables are part of the phase space, which differs from a perturbative approach around a fixed background.

Furthermore we need the integrated second order of the scalar constraint:
\ba\label{scalar2nd}
{}^{(2)}S(0) &=& \!\!\! \sum_{k}\! \kappa^{-1}
\Big(
2\beta E\, \epsilon^{cad}(\partial_a a_{bc}(k)-\partial_b a_{ac}(k)) {e^b}_d(-k) +
\beta^2 E^2\, ({a_a}^a(k) {a_b}^b(-k)- {a_a}^b(k) {a_b}^a(-k))+
\nn\\
&&\q\q\q
\beta^2 A^2\,( {e^a}_a(k) {e^b}_b(-k) - {e^a}_b(k) {e^b_a}(-k))+   
2 \beta^2 AE\,({a_a}^a(k) {e^b}_b(-k) +{a_a}^b(k) {e^a}_b(-k))
\Big) + 
\nn\\
&&\q
\gamma^{-1}\Big(
\tfrac{1}{2} \rho^2 +
\tfrac{1}{2} \,E^2\, \delta^{ab} (\partial_a\phi)(k) (\partial_b\phi)(-k) +
E^2 V'(\Phi)\, {e^a}_a(k)\,\phi(-k) + \nn\\
&&\q\q\q\q
\tfrac{1}{2}\, E^3 V''(\Phi)\phi(k)\phi(-k)+  \tfrac{1}{2}\,E V(\Phi)\,( {e^a}_a(k) {e^b}_b(-k) - {e^a}_b(k) {e^b_a}(-k) ) 
\Big) \q .
\ea

To construct complete observables we need to choose a set of clock variables $T^K(k)$ 
where $K\in \{0,Da,Ga;\,a=1,2,3\}$ and $k\in  2\pi\,\Zl^3$. 
This has to be done in a way such that at least the zeroth order of the matrix
\ba
\ca^K_j(k,k'):=\{ T^K(k) , C_j(k') \}
\ea
is invertible. This will ensure that at least to the
 lowest order the clock variables provide a good parametrization of the gauge orbits (and that is the reason we have to exclude the integrated diffeomorphism constraints, since these start at second order).

We will choose the clock $T^0(0)$ to have a non--vanishing zeroth order part and a vanishing first order part. All the other clocks should have vanishing zero order parts.
This will ensure that the new constraints $\tilde C_K(k)$ and the clocks $T^K$ have a similar structure as explained above equation (\ref{fassump}). (We are a bit more general here.)  The structure of the matrix $\ca$ is then as follows. The zeroth order of the matrix is of diagonal block form, that is
\ba\label{blockform}
{}^{(0)}\ca^{0}_j(0,k')&=&0 \q\q \text{if}\q j\neq 0 \q \text{or}\q k'\neq 0  \nn\\   
{}^{(0)}\ca^{K}_0(k,0)&=&0  \q\q   \text{if}\q K\neq 0 \q \text{or}\q k\neq 0     \q .
\ea
Since we will quite often need to exclude the index combinations $(j=0\,\,\text{and} \,\,k=0)$ as well as  $(K=0\,\,\text{and}\, \, k=0)$ from the set of indices to sum over, we will introduce indices $(\hat j,\hat K,\hat k)$ to signify that these do not assume the values $(j=0\, \text{and}\, k=0)$ or $(K=0\, \text{and}\,  k=0)$. (Also these indices do not include $j,K=Da$ and $k=0$, since we excluded the integrated diffeomorphism constraints.)  

Moreover the first order of the matrix element $\ca^{0}_0(0,0)=\{T^0(0),C_0(0)\}$ vanishes. This structure of the matrix $\ca$ ensures that the constraint 
\ba\label{dec11c}
\tilde C_0(0) &:=&\sum_{k'}  C_j(k') (\ca^{-1})^j_0(k',0)  \nn\\
             &=&  C_0(0)  (\ca^{-1})^0_0(0,0)\,\,   + \sum_{ \hat{k}' }  C_{\hat{j}}(\hat{k}') (\ca^{-1})^{\hat{j}}_0(\hat{k}',0)
\ea
has vanishing first order. To see this we have to convince ourselves that the first order of $(\ca^{-1})^0_0(0,0) $ and the zeroth order of $  (\ca^{-1})^{\hat j}_0(\hat{k}',0)$ are vanishing.  The latter follows from the fact that $\ca$ is of diagonal block form (\ref{blockform}), hence the inverse has the same kind of block form. Furthermore, the first order of the inverse of $\ca$ can be expanded as
\ba
{}^{(1)}(\ca^{-1})^j_0(k',0)  &=&
-\sum_{k'',k'''} {}^{(0)}(\ca^{-1})^j_K(k',k'')\, \,{}^{(1)}\ca^K_{m}(k'',k''')\,\,{}^{(0)}(\ca^{-1})^m_0(k''',0) 
\nn\\
&=&-\sum_{k''}  {}^{(0)}(\ca^{-1})^j_K(k',k'')\,\,{}^{(1)}\{T^K(k''),C_0(0)\} \,\,{}^{(0)}(\ca^{-1})^0_0(0,0)
\ea
For $j=0$ and $K=0$ the sum in the last line collapses to just one term
\ba
{}^{(1)}(\ca^{-1})^0_0(0,0)   
=- {}^{(0)}(\ca^{-1})^0_0(0,0)\,\,{}^{(1)}\{T^0(0),C_0(0)\}\,\,{}^{(0)}(\ca^{-1})^0_0(0,0) 
=0 \q 
\ea
where we used that the first order of $C_0(0)$ and $T^0(0)$ vanishes. \\ 

Hence the first order of the constraint $\tilde C_0(0)$ vanishes. Also this constraint is the sole one among the $\tilde C_K(k)$ with a non--vanishing zeroth order part:
\ba
{}^{(0)}\tilde C_K(k)=\sum_{k'}\,\,{}^{(0)}C_j(k')\,\,{}^{(0)}(\ca^{-1})^j_K(k',k)={}^{(0)}C_0(0)\,\,{}^{(0)}(\ca^{-1})^0_K(0,k)
\ea 
vanishes for $K\neq0$ or $k\neq0$ because of the block diagonal form of the zeroth order of $\ca$.

Let us consider the series for the complete observable $F_{[f;T^K]}(\tau)$ associated to a function $f$ and with parameter values $\tau^0(0)=\tau$ and $\tau^{\hat K}(\hat k)=0$:
\ba\label{nov221}
F_{[f;T^K]}(\tau) &\simeq&
\sum_{r=0}^\infty  \sum_{k_1,\ldots,k_r}
                \! \frac{1}{r!} \{ \cdots \{f,\tilde C_{K_1}(k_1)\},\cdots \} \tilde C_{K_r}(k_r) \}  
                  (\tau^{K_1}(k_1)-T^{K_1}(k_1)) \cdots  (\tau^{K_r}(k_r)-T^{K_r}(k_r))  
\nn\\
&\simeq & 
\sum_{r=0}^\infty  \sum_{s=0}^r \sum_{\hat k_1,\ldots, \hat k_r}  \frac{1}{(r-s)!s!} 
                   \{\cdots\{ f, \tilde C_0(0)\}_{(r-s)},\tilde C_{\hat K_1}(\hat k_1)\}, \cdots \}, \tilde C_{\hat K_s}(\hat k_s) \}  \times
\nn\\
&& \q\q\q\q\q\q\q
(\tau-T^0(0))^{r-s} \, \times \,
(-T^{\hat K_1}(\hat k_1)) \cdots  (-T^{\hat K_s}(\hat k_s))
\nn\\
&\simeq & 
\sum_{p=0}^\infty  \sum_{q=0}^\infty \sum_{\hat k_1,\ldots, \hat k_p}  \frac{1}{q!\, p!} 
                   \{\cdots\{ f, \tilde C_0(0)\}_{q}    (\tau-T^0(0))^q \,\,                          ,\,
\tilde C_{\hat K_1}(\hat k_1)\}, \cdots \}, \tilde C_{\hat K_p}(\hat k_p) \}      \times
\nn\\
&& \q\q\q\q\q\q\q
(-T^{\hat K_1}(\hat k_1)) \cdots  (-T^{\hat K_p}(\hat k_p))
\ea
where we denote by $\{\cdot,\cdot\}_q$ iterated Poisson brackets $\{f,g\}_q=\{\{f,g\}_{(q-1)},g\}$ and $\{f,g\}_0=f$. We used in the first step that we can arrange the constraints $\tilde C_K(k)$ in any order since they commute up to terms quadratic in the constraints. In the second step we exploited that $T^0(0)$ commutes with the constraints $\tilde C_{\hat K}(\hat k)$ up to terms proportional to the constraints. The result can be interpreted in the following way: The complete observable $F_{[f;T^K]}(\tau)$ can be calculated by first finding the complete observable corresponding to the single constraint $\tilde C_0(0)$ with parameter value $\tau$ and then computing the complete observable associated to this result with respect to the remaining constraints. One can also choose to perform the calculation in the other way around, i.e. first compute the complete observable with respect to the constraints $\tilde C_{\hat K}(\hat k)$ and then to deal with the constraint $\tilde C_0(0)$. Here one uses the fact that the clocks $T^{\hat K}(\hat k)$ commute with the constraint $\tilde C_0(0)$ up to terms proportional to the constraints.
\ba\label{nov222}
F_{[f;T^K]}(\tau) &\simeq&
\sum_{p=0}^\infty  \sum_{q=0}^\infty \sum_{\hat k_1,\ldots, \hat k_p}  \frac{1}{q!\, p!} 
                   \Big{\{}\{\cdots\{ f, \tilde C_{\hat K_1}(\hat k_1)\}, \cdots \} \tilde C_{\hat K_p}(\hat k_p) \} \times
\nn\\
&&\q\q\q\q\q\q\q\q\q\q\q\q
(- T^{\hat K_1} (\hat k_1) )  \cdots (-T^{\hat K_p}(\hat k_p))\, ,
\tilde C_0(0) \Big{\}}_{q}\,\, (\tau-T^0(0))^q   \q .\q\q
\ea

Assume that $f$ is a zeroth or first order quantity. Expanding the complete observable (\ref{nov221},\ref{nov222}) up to a certain order $m$ in the fluctuation variables, we see that we need the constraints $\tilde C_0(0)$ up to order $m$ for $f$ zeroth and up to order $(m+1)$ for $f$ first order. 
We need the remaining constraints up to order $m$ for $f$ first order and up to order $(m-1)$ for $f$ zeroth order 
. The reason for this is that the lowest order in $\{g, \tilde C_{\hat K}(\hat k)\}$ is $(n-1)$ if $g$ has lowest order $n$. However for each Poisson bracket with such a constraint the expression gets multiplied with the clock variable $T^{\hat K}(\hat k)$ which is at least of first order. On the other hand $\{g, \tilde C_{0}(0)\}$ is at least of order $n$ if $g$ has lowest order $n$, so one does not loose any order in the Poisson brackets with the constraint $\tilde C_0(0)$. Hence it is crucial that the constraint $\tilde C_0(0)$ has a vanishing first order part, otherwise a perturbational calculation in the usual sense is not possible.  

We can interpret expression (\ref{nov221}) in the following way: First we have to evolve the partial observable $f$ with respect to the constraint $\tilde C_0(0)$ which generates time evolution with respect to the clock $T^0(0)$. If we want to calculate the complete observable to a certain order $m$ we have to calculate this evolution up to terms of order $m$. This evolution can be broken up into ``free evolution'' described by the zeroth (for the zeroth order variables) and second order (for the first order variables) part of $\tilde C_0(0)$ and ``interaction processes'' described by the higher order parts (and the second order part for the interaction of the homogeneous variables with the inhomogeneities).

In a second step we have to calculate the gauge invariant extension of the (clock) time evolved function. This requires only a finite number of terms in the second (outer) sum since we can omit all terms with more than $m$ factors of the (inhomogeneous) clocks $T^{\hat K}(\hat k)$.

In the case that $f$ is of zeroth order, i.e. a ``background variable'', the zeroth order complete observable corresponds to the complete observable in the isotropic and homogeneous model. Higher (than first) order complete observables take into account the backreaction of the fluctuations onto the background.

For $f$ first order the first order complete observable describes (linearized) propagation of first order gauge invariant perturbations. The time evolution of these perturbations is expressed with respect to a physical clock (defined by a homogeneous variable). This is different from the usual theory of perturbations on a fixed background \cite{MFB}.


\section{ Transformation between different sets of clock variables}\label{trafo}

In this section we will explore the dependence of the complete variables on the choice of clock variables. Since one can understand the complete observables also as gauge invariant extensions of gauge restricted functions this will also enable us to connect different gauges.

 To derive a relation between the complete observables with respect to two different sets of clock variables $\{T^K(k)\}$ and $\{{T'}^K(k)\}$ we will come back to the interpretation of the complete observables: The complete observables $F_{[f;T^K]}(\tau^K)$ assigns to a phase space point $x$ the value of the function $f$ at the point $y$ on the gauge orbit through $x$ at which the clocks $T^K(k)$ coincide with the parameter values $\tau^K(k)$, that is $T^K(k)_{|y}=\tau^K(k)$.

If one replaces in $F_{[f;T^K]}(\tau^K)$ the $\tau^K(k)$ parameters by the complete observables $F_{[T^K(k);{T'}^L]}({\tau'}^L)$ one will get the value of $f$ at that point $z$ on the gauge orbit through $x$ at which the clocks $T^K(k)$ coincide with the complete observables $F_{[T^K(k);{T'}^L]}({\tau'}^L)$. Notice that whereas $T^K(k)$ changes along the gauge orbit the complete observable $F_{[T^K(k);{T'}^L]}({\tau'}^L)$ is constant along the gauge orbit, so with the assumption that the clocks provide a good parametrization of the gauge orbit the point $z$ is uniquely determined. Hence $z$ is a point on the gauge orbit through $x$ which has to satisfy 
\ba\label{nov223}
T^K(k)_{|z}\simeq F_{[T^K(k);{T'}^L]}({\tau'}^L)  \q . 
\ea
The complete observable $F_{[T^K(k);{T'}^L]}({\tau'}^L)$ on the right hand side gives the value of $T^K(k)$ on that point $y'$ on the gauge orbit through $x$ on which the clocks ${T'}^L(k)$ coincide with the parameter values ${\tau'}^L(k)$. Hence the point $z$ in (\ref{nov223}) has to coincide with the point $y'$, characterized by the condition ${T'}^L(k)_{|y'}={\tau'}^L(k)$. Therefore we can conclude that
\ba\label{nov224}
F_{[f;T^K]}(F_{[T^K(k);{T'}^L]}({\tau'}^L))\simeq F_{[f;{T'}^L]}({\tau'}^L)   \q .
\ea  
This gives us a relation between the complete observables with respect to two different sets of clock variables. If one wants to use this formula in order to obtain $F_{[f;{T'}^L]}({\tau'}^L)$ from $F_{[f;T^K]}(\tau^K)$ one needs the functional dependence of the complete observable $F_{[f;T^K]}$ on the parameter values $\tau^K(k)$. However with the exception of the parameter $\tau^0(0)$ we set these clock parameters to zero. One can nevertheless use formula (\ref{nov224}) if one Taylor expands the left hand side around some fixed values for the parameters $\tau^K(k)$. This would give a formula connecting complete observables with respect to clock variables ${T'}^K(k)$ with complete observables with respect to $T^K(k)$.

A simpler way to obtain a formula for $F_{[f;{T'}^L]}({\tau'}^L)$ as a function of complete observables with respect to the clocks $T^K(k)$ is to start with the power series for $F_{[f;{T'}^L]}({\tau'}^L)$ 
\ba\label{dec0701}
F_{[f;{T'}^L]}({\tau'}^L)
&\simeq&
\sum_{r=0}^\infty\sum_{k_1,\ldots,k_r}\frac{1}{r!}
\{\cdots\{ f,\tilde {C'}_{K_1}(k_1)\},\cdots\},\tilde {C'}_{K_r}\} \nn\\
&&
\quad\quad\quad\quad\quad ({\tau'}^{K_1}(k_1)-{T'}^{K_1}(k_1))\cdots ({\tau'}^{K_r}(k_r)-{T'}^{K_r}(k_r))
\ea
where $\tilde{C'}_K(k)=\sum_k C_j(k')({A'}^{-1})^j_K(k',k)$ and ${A'}^K_j(k,k')=\{{T'}^K(k),C_j(k')\}$. Now we take both sides of equation (\ref{dec0701}) as ``input function'' $f$ for the complete observable $F_{[f,T^K]}(\tau^K)$. The left hand side does not change, since the complete observable associated to a Dirac observable is given by the Dirac observable itself. Therefore we get
\ba\label{dec0702} 
F_{[f;{T'}^L]}({\tau'}^L)
&\simeq&
\sum_{r=0}^\infty\sum_{k_1,\ldots,k_r}\frac{1}{r!}
F_{[  \{\cdots\{ f,\tilde {C'}_{K_1}(k_1)\},\cdots\},\tilde {C'}_{K_r}\}            \,\, ;\,T^K ]}(\tau^K)
\nn\\
&&
\quad\quad\quad\quad\quad 
({\tau'}^{K_1}(k_1)-F_{[{T'}^{K_1}(k_1)\,; \, T^K]}(\tau^K)   )\cdots ({\tau'}^{K_r}(k_r)- F_{[ {T'}^{K_r}(k_r) \,; \, T^K]}(\tau^K)    ) \quad \quad
\ea
as a formula for a complete observable with respect to clocks ${T'}^L$ as a function of complete observables with respect to clocks ${T^K}$.

In a first order approximation (for $f$ first order) and for the case that the clocks $T^0(0)={T'}(0)$ and the corresponding parameter values $\tau^0(0)={\tau'}^0(0)=\tau$ coincide formula (\ref{dec0702}) reduces to
\ba\label{aaaa}
{}^{(1)}F_{[f;{T'}^L]}({\tau})
&\simeq&
{}^{(1)}F_{[f;{T}^K]}({\tau})-
\sum_{\hat k} 
\, {}^{(0)}F_{[{}^{(0)}\{f,\tilde {C'}_{\hat K}(\hat k)\}\,;T^L]}(\tau)
\,\,{}^{(1)}F_{[{T'}^{\hat K}(\hat k)\,;T^L]}(\tau)  \quad .
\ea 
Here we set all the other parameter values $\tau^{\hat K}(\hat k)={\tau'}^{\hat K}(\hat k)=0$ to zero and we used that $\tau=F_{[{T'}^0(0)\,;\, T^L]}(\tau)$. 

If we have found the first order complete observables associated to some basis of phase space functions with respect to one set of clocks $T^{\hat K}(\hat k)$, we can calculate the first order complete observables with respect to another set of clocks with the help of formula (\ref{aaaa}). Formulae for higher order complete observables can be derived by expanding (\ref{dec0702}) to the appropriate order.

\section{Lapse and shift functions}\label{lapseandshift}

As explained earlier the complete observables $F_{[f;T^k]}(\tau^K)$ can also be understood as gauge invariant extensions of the phase space function $f$ using the gauge $\{T^K(k) =\tau^K(k)\,\,\forall K,k\}$. Considering the complete observables just for one fixed set of parameters $\tau^K$ would correspond to a ``frozen time'' picture. The time evolution is generated by a constraint, that is time evolution is a gauge transformation. Fixing all gauge degrees of freedom would also mean to consider a fixed time. However, we can choose a one--parameter family of gauge fixings, as for instance $T^0(0)=\tau,\, T^{\hat K}(\hat k)=0$ for $\tau$ in (some subset of) $\Rl$, that would represent a varying time. 

A phase space point on the constraint hypersurface gives rise to a solution of the equation of motion, that is a space--time manifold. The one--parameter family of gauge fixings defines a foliation of this space--time manifold, as well as spatial coordinates on each of the leafs of the foliation. Hence we can find (phase space dependent) lapse functions and shift vectors using this foliation and characterize our choice for the clock variables and the one--parameter family of $\tau$--parameters. 

Lapse and shift can be used to construct the four--dimensional metric using the four--dimensional coordinates given by the foliation. This allows one to compare the results of this approach to (covariant) methods utilizing gauge fixing. 

From (\ref{feb036}) we see that the (gauge) generator for a translation in the $\tau^0(0)$ parameter is given by $\tilde C_0(0)$. Hence the lapse function\footnote{
Note that we are using the scalar constraint $S$ here, which has density weight 2. Usually one defines the lapse function $N^\perp$ as the coefficient in front of the Hamiltonian constraint $C_\perp=(\sqrt{q})^{-1}C$, which has density weight one. Hence we have $\cn^\perp=\sqrt{q}\cn^0$.
}  
$\cn^0$ and shift vector $\cn^{Da}$ can be read off as the coefficients in front of the scalar and diffeomorphism constraint. We will also define $\cn^{Ga}$ as the coefficient in front of the Gauss constraint. 

This motivates the definition
\ba\label{nov241}
\cn^j(k):=(\ca^{-1})^j_0(-k,0) , \q ,
\ea
so that we can write
\ba
\tilde C_0(0)=\sum_{k} \cn^j(k)C_j(-k)
\ea
for our ``time evolution'' generator $\tilde C_0(0)$. Note that if one uses a gauge fixing one would restrict the consideration to the gauge fixing hypersurface $\{T^K(k)=\tau^K(k)\}$, so if one compares (\ref{nov241}) to gauge fixing one should omit all terms vanishing on this hypersurface.


\section{Clock variables and Hamiltonians}  \label{clocks}

The clock variables should be chosen such that the zeroth order of the matrix $\ca^K_j(k,k')$ is invertible. Once one has found such clock variables one can in principle define new clock variables by multiplying the old clock variables with the zeroth order of the inverse matrix $(\ca^{-1})^K_j(k,k')$. These new clock variables will lead to a new matrix, the zeroth order of which will be given by the identity matrix (with the exception of the entry $\ca^0_0(0,0)$ which might differ from $1$).   

However these new clock variables might be not very convenient to deal with, since also the first order clocks $T^{\hat K}(\hat k)$ will depend on the homogeneous variables, leading to additional terms for the higher orders of the matrix $A^K_j(k,k')$  coming from the Poisson bracket between the homogeneous variables in the clock variables and the constraints.

Here we will specify the ``inhomogeneous'' clock variables $T^{\hat K}(\hat k)$. Hence we allow zero momentum $k$ for the Gauss clock $T^{Ga}$ but not for the scalar clock $T^0$ and not for the diffeomorphism clock $T^{Da}$ (because of the linearization instabilities).
One choice for the clock variables is
\ba\label{choiceD}
T^{Ga} &=&\epsilon^{abc}e_{bc}  \nn\\
          &=&\epsilon^{abc}({}^{AT}e_{bc}+{}^{LT}e_{bc}+{}^{TL}e_{bc}) \nn\\
T^{Da} &=& -W^{-2} ( -\tfrac{1}{2} W^{-2}\partial^a\partial_d\partial_e +\tfrac{1}{2} \partial^a \delta_{de} -\partial_e \delta^a_d -\partial_d \delta^a_e)e^{de} \nn\\
&=& 
-W^{-2} ( -\partial^a\,{}^{LL}{e^d}_d+\frac{1}{2}\partial^a\,{}^{T}{e^d}_d -\partial_e\,{}^{TL}e^{ae}-\partial_d\,{}^{LT}e^{da})  \nn\\
T^0 &=& -W^{-2}\partial_a\partial_b e^{ab} \nn\\
        &=& {}^{LL}{e^a}_a
\ea
where $W:=\sqrt{-\partial^e\partial_e}$. Here we introduced a tensor mode decomposition for the gravitational variables (with respect to the flat metric $\delta_{ab}$), the notation is explained in appendix \ref{tensor modes}.

This set of clock variables is obviously Abelian and leads to the following zeroth order for the Poisson brackets between the clocks and the constraints:
\begin{xalignat}{3}\label{AchoiceD}
&\{T^{Ga}(k),{}^{(1)}\!G_b(k')\}=\!2\beta E \delta^a_b { \delta_{k,-k'} } 
\!\!\!\!\!
&&\{T^{Ga}(k),{}^{(1)}\!D_b(k')\}=\!E \partial'_c {\epsilon_b}^{ca}\delta_{k,-k'}  
\!\!\!\!\!
&&\{T^{Ga}(k),{}^{(1)}\!S(k')\}=0    \nn\\
&\{T^{Da}(k),{}^{(1)}\!G_b(k')\}=0
&&\{T^{Da}(k),{}^{(1)}\!D_b(k')\}=E \delta^a_b \delta_{k,-k'} 
&&\{T^{Ga}(k),{}^{(1)}\!S(k')\}=0    \nn\\
&\{T^0(k),{}^{(1)}\!G_b(k')\}=0 
&&\{T^0(k),{}^{(1)}\!D_b(k')\}=  0 
&&\{T^0(k), {}^{(1)}\!S(k')\}=\!-4\beta^2AE^2   \delta_{k,-k'}
\end{xalignat}
By adding to the Gauss clock a term proportional to the diffeomorphism clock we can define new clock variables which lead to a diagonal (except for the $\delta_{k,-k})$ factor) matrix ${}^{(0)}A^{{K}}_j(k,k')$.
\ba\label{choiceE}
 T^{Ga} &=&\epsilon^{abc} e_{bc} - {\epsilon^{ab}}_c\partial_b\, {}^{D}T^C=
\epsilon^{abc} (  e_{bc} -W^{-2}\partial_b\partial_e {e_c}^{e}-W^{-2}\partial_b\partial_e {e^e}_c )
   \nn\\
          &=&   \epsilon^{abc} {}^{(AT+LT+TL)}\!e_{bc}-\epsilon^{abc}(W^{-2}\partial_b\partial^e\,{}^{TL}\!e_{ce}+W^{-2}\partial_b\partial^d \,{}^{LT}\!e_{dc})  \q        
\ea
where we assume $k \neq 0$. (The $T^{Ga}(k=0)$ clock is still given by $\epsilon^{abc}e_{bc}$.)

With this new Gauss clock we have
\begin{xalignat}{3}\label{AchoiceE}
&\{T^{Ga}(k),{}^{(1)}\! G_b(k')\}=\! 2\beta E \delta^a_b \delta_{k,-k'}
\!\!\!\!\!  
&&\{T^{Ga}(k),{}^{(1)}\! D_b(k')\}=0  
&&\{T^{Ga}(k),{}^{(1)}\! S(k')\}=0    \nn\\
&\{T^{Da}(k),{}^{(1)}\! G_b(k')\}=0
&&\{T^{Da}(k),{}^{(1)}\! D_b(k')\}=\!E \delta^a_b \delta_{k,-k'} 
\!\!\!\!\!
&&\{T^{Da}(k),{}^{(1)}\! S(k')\}=0    \nn\\
&\{T^0(k),{}^{(1)}\! G_b(k')\}=0 
&&\{T^0(k),{}^{(1)}\! D_b(k')\}=  0 
&&\{T^0(k), {}^{(1)}\! S(k')\}=\!-4\beta^2AE^2   \delta_{k,-k'}
\end{xalignat}
for the zeroth order of the Poisson brackets between the clocks and the constraints. As one can see from formula (\ref{dec0702}), which gives the relation between complete observables using different sets of clock variables, the complete observables using either the clocks (\ref{choiceD}) or the clocks (\ref{choiceE}) coincide (for parameter values $\tau^{\hat K}(\hat k)=0$). The reason for this is, that both sets of clocks define the same gauge fixing surface.


Assume that one has chosen a clock $T^0(0)$, as for instance $T^0(0)=\Phi$. Then we can define gauge invariant functions, that generate the evolution for the complete observables in the $\tau=\tau^0(0)$ parameter. According to section \ref{appdirac} we have to find a momentum $P_0$ conjugated to $T^0(0)$, which has to commute (weakly) with the other clocks $T^{\hat K}(\hat k)$. For the example above we could choose $P_0=\Pi$. In general we will assume that $P_0$ has non--vanishing zeroth order part and vanishing first order part. 

(Since we are only interested in the generator for the evolution in $\tau^0(0)$, we do not have to specify momenta conjugated to the clocks $T^{\hat K}(\hat k)$. However a natural choice would be to choose the first order of the constraints ${}^{(1)}\tilde C_{\hat K}(\hat k)$. For issues arising because of the linearization instabilities see appendix \ref{lininstab}.)  

Then we can define the physical Hamiltonian as $H_0(\tau):=F_{[h_0\, , T^K]}(\tau)$ where
\ba
h_0=-P_0 \simeq -P_0+\tilde C_0(0)  \q .
\ea
With this physical Hamiltonian we can write
\ba\label{moon1}
\frac{\bd }{\bd \tau} F_{[f\, ,T^K]}(\tau)=\{F_{[f\, ,T^K]}(\tau), H_0(\tau)\} 
\ea
for functions $f$ that Poisson commute with $P_0$ and the clocks $T^K$. If $f$ does not commute with $P_0$ we have to add a term $F_{[\{f,P_0\}\, , \, T^K]}(\tau)$ to the right hand side of equation (\ref{moon1}). This term has the same purpose as the additional time derivative $\partial_t f$ that appears in explicit time dependent Hamiltonian systems in classical mechanics, where the time evolution equations are given by
\ba
\frac{\bd }{\bd t} F =\{F ,H\}+\frac{\partial}{\partial t}f   \q .
\ea
In general the Hamiltonian $H_0(\tau)$ will have a non--vanishing zeroth order part (as long as one does not choose ${}^{(0)}\! P_0={}^{(0)}\! \tilde C_0(0)$) and a vanishing first order part. If one interprets $H_0(\tau)$ as an energy, this shows that also the zeroth order variables contribute to this energy.  $H_0(\tau)$ will be $\tau$-- independent (i.e. energy is conserved) if $P_0$ is a gauge invariant function. This would be the case for $P_0=\Pi$ and constant potential for the scalar field.

Another choice for the clock variables, which is, as we will see in appendix \ref{longo}, related to the so called longitudinal gauge \cite{MFB} is given by
\ba\label{choiceLong}
T^{Ga} &=&\epsilon^{abc}e_{bc}  \nn\\
          &=&\epsilon^{abc}({}^{AT}e_{bc}+{}^{LT}e_{bc}+{}^{TL}e_{bc}) \nn\\
T^{Da} &=& -W^{-2} ( -\tfrac{1}{2} W^{-2}\partial^a\partial_d\partial_e +\tfrac{1}{2} \partial^a \delta_{de} -\partial_e \delta^a_d -\partial_d \delta^a_e)e^{de} \nn\\
&=& 
-W^{-2} ( -\partial^a\,{}^{LL}{e^d}_d+\frac{1}{2}\partial^a\,{}^{T}{e^d}_d -\partial_e\,{}^{TL}e^{ae}-\partial_d\,{}^{LT}e^{da})  \nn\\
T^0 &=&       W^{-2}(\tfrac{1}{2} \delta^{cd} +\tfrac{3}{2}W^{-2}\partial^c\partial^d)a_{cd}                                                        \nn\\
&=&W^{-2}(\tfrac{1}{2}\,{}^{T}{a^d}_d-\,{}^{LL}{a^d}_d)  \q .
\ea 
This set of clocks differs from (\ref{choiceD}) only in the scalar clocks. Notice that the scalar clocks and the diffeomorphism clocks do not commute. The zeroth order parts of the Poisson brackets between constraints and clocks are given by
\begin{xalignat}{3}\label{AchoiceLong}
&\{T^{Ga}(k),{}^{(1)}\! G_b(k')\}=2\beta E \delta^a_b \delta_{k,-k'}  
\!\!\!
&&\{T^{Ga}(k),{}^{(1)}\! D_b(k')\}=E \partial'_c {\epsilon_b}^{ca}\delta_{k,-k'}  
\!\!\!\!\!\!\!
&&\{T^{Ga}(k),{}^{(1)}\! S(k')\}=0    \nn\\
&\{T^{Da}(k),{}^{(1)}\! G_b(k')\}=0
&&\{T^{Da}(k),{}^{(1)}\! D_b(k')\}=E \delta^a_b \delta_{k,-k'} 
\!\!\!
&&\{T^{Ga}(k),{}^{(1)}\! S(k')\}=0    \nn\\
&\{ T^0(k),{}^{(1)}\! G_b(k')\}=W^{-2} \partial_{b}\delta_{k,-k'}
&&\{T^0(k),{}^{(1)}\! D_b(k')\}= -W^{-2}A\partial_b  \delta_{k,-k'}
&&\{T^0(k), {}^{(1)}\! S(k')\}=2E   \delta_{k,-k'}\, .
\end{xalignat}

For these clocks, which are related to the longitudinal gauge, we cannot give a physical Hamiltonian along the lines of section \ref{appdirac}. 
The reason for this is, that the clocks do not Poisson commute with each other. However according to equation (\ref{cham3}), which gives the Dirac bracket, the term that arises because of the Non--Abelianess of the clocks is 
\ba
\sum_{k,k'} \{ f, \tilde C_{K}(k)\}\{T^K(k),T^M(k')\}\{\tilde C_M(k'),h_0\} \q .
\ea
This term is at least of second order if $f$ and $h_0$ are gauge invariants to first order. Under these conditions a physical Hamiltonian valid for the linearized theory can be defined in the same way as for the Abelian clocks (\ref{choiceD}).

\subsection{Scalar fields as clocks} \label{scalar}

The clock variables we introduced so far are quite non--local, i.e. they require for their definition inverse derivatives (in the form of $W^{-2}=|k|^{-2}$) or the inverse of the matrix $A^K_j$ requires inverse derivatives. One way to avoid this, is to use scalar fields as clocks. Scalar fields are used quite often as clocks, see for example \cite{kuchar4, smolincosmo, bd2, brand, hartle1, phantomtt} . As explained in \cite{bd2} using scalar fields as clocks can lead to huge simplifications for the calculation of complete observables. This is related to the fact that scalar fields provide a local characterization of spacetime points (as opposed to for instance the longitudinal modes of the metric fields, which rather characterize a foliation of spacetime).

On the one hand we are interested in a homogeneous background, on the other hand we want to use the values of four scalar fields to define a coordinate system, i.e. at least the scalar fields defining the spatial coordinates have to be non--homogeneous. Hence we will choose these scalar fields to be massless, that is having vanishing potential. Since we assumed the spatial manifold $\Sigma$ to have topology $S^1 \times S^1 \times S^1$ we will assume that the scalar fields $\varphi^A,\,A=1,2,3$ take values in $S^1$ (parametrized by the values of the interval $[0,1[$).

We will use another scalar field $\varphi^0$, taking values in $\Rl$, as the time coordinate, that is the clock for the scalar constraint. The division into background variables and perturbations is now  
\begin{xalignat}{2}\label{dec1001}
&{A_a}^j(\sigma)=A\,\beta \delta_a^j+{a_a}^b(\sigma)\,\beta\delta^j_b \q ,\q\q 
&&{E^a}_j(\sigma)=E\,\beta^{-1}\delta^a_j +{e^a}_b(\sigma)\,\beta^{-1} \delta^b_j  \nn\\
&\varphi^0(\sigma)=\Phi^0+\phi^0(\sigma) \q, \q\q  
&&\pi_0(\sigma)=\Pi_0 +\rho_0(\sigma)  \nn\\
&\varphi^A(\sigma)=\delta^A_a\sigma^a+\phi^A(\sigma)      \q, \q\q  
&&\pi_A(\sigma)=0+\rho_A(\sigma)           
\end{xalignat}
where $\pi_0,\pi_A$ are the momenta conjugate to the scalar fields $\varphi^0,\varphi^A$, respectively. The integrals of $\phi^0,\rho^0$ over the spatial manifold are constrained to vanish, this is not the case for the variables $\phi_A,\rho_A$. (For the fields $\varphi^A,\pi_A$ we perturb around a fixed value and not around their averaged value.) 
 

The Poisson brackets between the gravitational variables are as before (\ref{poisfeb}), for the matter fields we have
\begin{xalignat}{3}\label{dec1001a}
& \{\Phi^0,\Pi_0\}=
\gamma_0
,&& \{\phi^0(\sigma),\rho_0(\sigma')\}=\gamma_0\delta(\sigma,\sigma')-
\gamma_0
,&& \{\phi^A(\sigma),\rho_A(\sigma')\}=\gamma_A\delta(\sigma,\sigma')
\end{xalignat}
where $\gamma_0,\gamma_A$ are the coupling constants for the scalar fields $\varphi^0,\varphi^A$ respectively. The matter parts of the diffeomorphism and scalar constraints are given by  
\ba
{}_{mat}D_a &=&\gamma_0^{-1}\pi_0\partial_a\varphi^0+\sum_{A=1,2,3}\gamma_A^{-1}\pi_A\partial_a\varphi^A 
\nn\\
&=& \gamma_0^{-1}\Pi_0\partial_a\phi^0 +\sum_{A=1,2,3}\gamma_A^{-1}\rho_A \delta^A_a+ O(2)
\nn\\
{}_{mat}S &=&\frac{1}{2}\gamma_0^{-1}(\pi_0^2 +qq^{ab}\partial_a\varphi^0\partial_b\varphi^0+ 2q V_0(\varphi^0))
+\frac{1}{2}\sum_A \gamma_A^{-1}(\pi_A^2 +qq^{ab}\partial_a\varphi^A\partial_b\varphi^A) 
\nn\\
&=& \gamma_0^{-1}(\frac{1}{2}\Pi_0^2+E^3 V_0(\Phi^0))+\frac{1}{2}\sum_A \gamma_A^{-1} E^2 + \gamma_0^{-1}(\Pi_0\rho_0+E^3V_0'(\Phi^0)\phi^0+E^2V_0(\Phi^0){e^a}_a) \, +
\nn\\
&&\q \sum_A\gamma_A^{-1}( e^{ab}\delta^A_a\delta^A_b+E^2\delta^{Aa}\partial_a\phi^A    )+O(2) \q .
\ea
We could add more scalar fields, these will have the same kind of contribution as the scalar field $\varphi^0$. Note that the first order of the integrated diffeomorphism constraint does not vanish anymore
\ba
{}^{(1)}D_a(k=0)=
\int_\Sigma {}^{(1)}D_a(\sigma)\, \bd \sigma =
\int_\Sigma \sum_{A=1,2,3}\gamma_A^{-1}\rho_A \delta^A_a\, \bd \sigma
\ea
so the problem of the linearization instabilities does not occur for this choice of background (because the background values of the scalar fields break the translational symmetry). 

The first order of the integrated scalar constraint is still vanishing, showing that there actually exists an exact solution of the equation of motions where all the perturbation variables vanish identically for all times.   

Now we can choose as clock variables
\begin{xalignat}{3}
T^{Ga}(k) =\epsilon^{abc}e_{bc}(k)
&&T^{Da}(k) = \sum_A \varphi^A(k) \delta_A^{a} 
&&T^0(k) =\varphi^0  (k)     \q 
\end{xalignat}
for all wave vectors $k=0$ and $k\neq 0$. As parameter values one has to choose $\tau^{Ga}(k)=0,\,\,\tau^{Da}(k)=
\int_\Sigma \exp(-ik\sigma)\sigma^a \bd \sigma$ and $\tau^0(k)=0$ for $k\neq 0$ as well as $\tau^0(0)=\tau$. The zeroth order of the Poisson bracket between the Gauss clock and the constraints is as before (\ref{AchoiceD}), for the Poisson brackets between the diffeomorphism and scalar clock and the constraints we have
\ba
{}^{(0)}\{T^{Da}(k),D_b(k')\}=\delta^a_b\delta_{k,-k'} \q , \q\q\q {}^{(0)}\{T^0(k),S(k')\}=\Pi_0\delta_{k,-k'}
\ea
with all the other (zeroth order) Poisson brackets vanishing. Hence the zeroth order of the matrix $\ca^K_j(k,k')$ is invertible on phase space points where $\Pi_0 \neq 0$ and $E\neq 0$ (from the commutator of the Gauss clock with the Gauss constraint). 

These considerations show that one can apply the perturbative formalism also if one uses scalar fields as clock variables. Moreover the problem of linearization instabilities does not occur.


\section{First order perturbations: scalar modes}\label{scalarmodes}

Let us consider the first order of the complete observables in more detail. We will assume that we are dealing with one scalar field and use the gravitational fields to define the non--homogeneous clock variables. Starting from the partial differential equation (\ref{feb036}) for the complete observables we will derive the equations of motion for the scalar mode perturbations (to first order). The equations of motions for the tensor modes are derived in appendix \ref{gravitons}. Consider the ``time evolution'' equation 
\ba
\frac{\bd}{\bd\tau} F_{[f;T^K]}(\tau)=F_{[\{f,\tilde C_0(0)\}\,;T^K]}(\tau)
\ea
where $\tau=\tau^0(0)$ is the parameter associated to the clock ${}^{S}T(k=0)$. To simplify the formulae we will introduce the notation $[[f]](\tau):= F_{[f;T^K]}(\tau)$ and suppress the dependence from the choice of clock variables. 

Hence we have
\ba\label{co1}
\frac{\bd}{\bd \tau} [[f]](\tau) \simeq [[\{f, \tilde C_0(0)\}]](\tau) \simeq \sum_{k'} \,  [[\{f,C_j(k')\}]](\tau)\,[[ (\ca^{-1})^j_0(k',0)]](\tau)
\ea
as the differential equation satisfied by the complete observable associated to $f$. 
With the introduction of lapse and shift functions (see section \ref{lapseandshift})
\ba 
 \cn^j(k):= (\ca^{-1})^j_0(-k,0)  
\ea
we obtain the following system of differential equations 
\ba\label{co2}
\frac{\bd}{\bd \tau} {}^{(1)}[[\phi(k)]](\tau) & \simeq &
  {}^{(0)}[[\cn^0(0)]](\tau)\,\, {}^{(1)}[[\rho(k)]](\tau) +{}^{(1)}[[\cn^0(k)]](\tau)\,\,{}^{(0)}[[\Pi]](\tau) 
  \\
\frac{\bd}{\bd \tau} {}^{(1)}[[\rho(k)]](\tau)& \simeq &
 {}^{(0)}[[\cn^0(0)]](\tau)\,\,  {}^{(1)}[[ E^2\, (\partial^a\partial_a \phi)(k)-E^2 V'(\Phi)\, {e^a}_a(k)-E^3 V''(\Phi)\, \phi(k)]](\tau)
-  \nn\\
&&
 {}^{(1)}[[\cn^0(k)]] (\tau)\,\,{}^{(0)}[[E^3V'(\Phi)]](\tau) -{}^{(1)}[[\partial_a \cn^{Da}(k)]] (\tau)\,\,{}^{(0)}[[ \Pi]](\tau)
\ea
for the first order complete observables associated to the fluctuations $\phi(k),\rho(k)$ in the scalar field and its conjugate momentum. Here we assume $k\neq 0$.

Now on the right hand side of the equations in (\ref{co2}) there appear also functions of the gravitational degrees of freedom, so in principle one would have to add differential equations for these gravitational degrees of freedom. However, if we are dealing with only one scalar field in our system, there should be only one unconstrained scalar mode degree of freedom. Indeed we can use 
\ba
[[C_j(k)]](\tau) \simeq 0 \q ,\q\q  [[T^{\hat K}(\hat{k})]](\tau)\simeq 0
\ea
to express the (first order complete observables associated to the) lapse and shift functions as well as ${e^a}_a$ as functions of the (first order complete observables associated to the) scalar field $\phi$ and its conjugated momentum $\rho$ and the homogeneous variables. 

For instance for the choice (\ref{choiceD}) of clock variables we have
\ba\label{co4}
\partial_a \, T^{Ga}  &=& \epsilon_{abc}\partial^a\,{}^{AT}e^{bc}\\
\partial_a\,T^{Da} &=&  -W^{-2}(-\tfrac{3}{2}\partial_a\partial_b\,{}^{LL}e^{ab} +\tfrac{1}{2}\partial^d\partial_d \delta_{ab}
({}^{LL}e^{ab}+{}^{T}e^{ab})) \nn\\
&=&  (- {}^{LL}{e^a}_a +\tfrac{1}{2}\,{}^{T}{e^a}_a )    \\
T^0&=&  {}^{LL}{e^a}_a  \q .
\ea
Hence we can use
\ba \label{dec1101}
[[{}^{AT}e^{ab}(k)]]\simeq 0 \q,\q\q [[{}^{T}e^{ab}(k)]]\simeq 0 \q ,\q\q [[{}^{LL}e^{ab}(k)]]\simeq 0  \q .
\ea
Furthermore using the first order of the constraints (\ref{cosmo4}) and the relation (\ref{dec1101}) gives us
\ba \label{dec1102} 
&& [[{}^{AT}a^{bc}(k) ]]  \simeq  0+O(2)  \q , \q\q
 [[ {}^{T} {a^b}_b(k) ]]  \simeq  -\frac{\kappa}{\gamma}\,E^{-1}\Pi \, \phi(k)+ O(2)  \nn \\
&& [[{}^{LL}{a^b}_b(k)]]  \simeq   -\frac{\kappa}{\gamma}\,(4\beta^2 A E^2)^{-1} (\Pi\,\rho(k)+E^3V'(\Phi)\,\phi(k))+\frac{\kappa}{\gamma}E^{-1}\Pi\,\phi(k) +O(2)  \q .
\ea
This shows that we can express the gravitational scalar modes as a combination of the matter scalar modes. 

The lapse and shift functions (for $k\neq 0$) for the clock variables (\ref{choiceD}) can be found to be (here we abbreviate $\cn:={}^{(0)}\cn^0(0)$) 
\ba\label{dec1103}
{}^{(1)}\cn^{Da}(k)&=&-\cn\sum_{k'}{}^{(0)}\!(\ca^{-1})^{Da}_K(-k,k')\,\,{}^{(1)}\!\ca^K_0(k',0)
= -\cn\sum_{k'} E^{-1}\delta_{-k',-k}\{ T^{Da}(k'), {}^{(2)}S(0)\} \nn\\
&=&- \cn\,E\, 2\beta^2 W^{-2}(\partial^a\,{}^{LL}{a^b}_b-\tfrac{1}{2}\partial^a\,{}^{T}{a^b}_b) \,+ O(T)
\nn\\
{}^{(1)}\cn^0(k) &=& -\cn \sum_{k'}{}^{(0)}\!(\ca^{-1})^{0}_K(-k,k')\,\,{}^{(1)}\!\ca^K_0(k',0)
=-\cn\sum_{k'}(-4\beta^2AE^2)^{-1}\delta_{-k,-k'}\{T^0(k'),  {}^{(2)}S(0)\} \nn\\
&=& -\cn \tfrac{1}{2}\, A^{-1}\,\,{}^{T}{a^b}_b \,+O(T)
\ea
where $O(T)$ denotes terms which vanish with $T^{\hat K}(\hat k)$. Using the equations (\ref{dec1102}) the differential equations for the scalar matter field becomes
\ba\label{dec1104}
\frac{\bd}{\bd \tau} {}^{(1)}[[\phi(k)]](\tau) & \simeq & {}^{(1)}[[\cn (\rho(k)+\tfrac{1}{2}\tfrac{\kappa}{\gamma}(AE)^{-1}\Pi^2\phi(k))]](\tau)  \nn\\
\frac{\bd}{\bd \tau} {}^{(1)}[[\rho(k)]](\tau) & \simeq & {}^{(1)}[[
\cn (E^2\partial^a\partial_a\,\phi(k)-3\tfrac{\kappa}{\gamma}\beta^2\Pi^2\phi(k)-E^3\,V''(\Phi)\phi(k)+\tfrac{1}{2}\tfrac{\kappa}{\gamma}(AE)^{-1}\Pi^2\, \rho(k))]](\tau)  \q . \nn\\
\ea
These equations have to be supplemented with the differential equation for the homogeneous variables $H=A,E,\Pi$ or $\Phi$, here it is sufficient to consider the zeroth order of this equation:
\ba
\frac{\bd}{\bd \tau}{}^{(0)}[[H]](\tau )\simeq {}^{(0)}[[\cn \, \{H, S(0)\}]]  \q .
\ea

Assume that one can find the general solution for these differential equations in dependence on initial data for some parameter value $\tau=\tau_0$. Since the differential equations (\ref{dec1104}) is linear in the fluctuation fields such a solution for instance for the scalar field can be written as
\ba
{}^{(1)}\![[\phi(k)]](\tau) =
G_1\big( (\tau-\tau_0);k; {}^{(0)}\! [[H]](\tau_0)\big)\,\,{}^{(1)}\![[\phi(k)]](\tau_0)
                          \,+ \,
G_2\big( (\tau-\tau_0);k; {}^{(0)}\! [[H]](\tau_0)\big)\,\,{}^{(1)}\![[\rho(k)]](\tau_0) \nn\\ {}\!\!\!
\ea
where $G_1,G_2$ can be understood as generalized (free) propagator functions. Now the complete observables restricted to the gauge fixing surface have to satisfy
\ba
{}[[\phi(k)]](\tau)_{|T^0(0) =\tau,\, T^{\hat K}(\hat k)=0}\, &\simeq&  \phi(k) \nn\\
{} [[H]](\tau)_{|T^0(0)  =\tau, \,T^{\hat K}(\hat k)=0} \, &\simeq&  H  \q .
\ea
For the complete observable associated to the scalar field we therefore have
\ba\label{dec11a}
{}^{(1)}[[\phi(k)]](\tau)_{|\, T^{\hat K}(\hat k)=0  }  \simeq 
G_1\left( (\tau-T^0(0));\,k;\, H\right)\,\,\phi(k)
                         \, +\, 
G_2\left( (\tau-T^0(0));\,k;\, H\right)\,\,\rho(k) \q .
\ea
Now we only have to determine ${}^{(1)}[[\phi(k)]](\tau)$ away from the hypersurface $\{ T^{\hat K}(\hat k)=0  \}$. This is done by replacing $\phi(k)$ and $\rho(k)$ in (\ref{dec11a}) by their first order gauge invariant extensions in $T^{\hat K}(\hat k)$--direction
\ba
\phi(k)\,  &\rightarrow & \, \phi(k)-\sum_{\hat k} {}^{(0)}\{ \phi(k)\,,\, \tilde C_{\hat K}(\hat k) \}\,\, T^{\hat K}(\hat k)\nn\\
\rho(k)\,  &\rightarrow &\, \rho(k)-\sum_{\hat k}{}^{(0)}\{\rho(k),\tilde C_{\hat K}(\hat k)\}\,\, T^{\hat K}(\hat k) \q .
\ea
Hence the first order complete observable associated to the matter scalar mode $\phi(k)$ is given by
\ba
{}^{(1)}[[\phi(k)]](\tau)  &\simeq &
G_1\left( (\tau-T^0(0));\,k;\, H\right)\,\,\big(\,\,\phi(k)-\sum_{\hat k} {}^{(0)}\{ \phi(k)\,,\, \tilde C_{\hat K}(\hat k) \}\,\, T^{\hat K}(\hat k) \,\,\big)
                         \, +\,  \nn\\
&& G_2\left( (\tau-T^0(0));\,k;\, H \right)\,\,\big( \,\,\rho(k)-\sum_{\hat k}{}^{(0)}\{\rho(k),\tilde C_{\hat K}(\hat k)\}\,\, T^{\hat K}(\hat k)\,\,\big) \q .
\ea

The first order complete observables can be understood to describe the ``free'' propagation of the perturbations on the cosmological background. Higher order complete observables will take care of interaction processes\footnote{
See \cite{bd3} for an explicit example of a complete observable taking into account interaction processes in a second order approximation around flat space.
} given by the higher (than second) order parts of the constraint $\tilde C_0(0)$  as well as of backreaction terms arising from the coupling of homogeneous and inhomogeneous variables in the constraint. In the next section we will consider second order complete observables associated to a zeroth order phase space function, which capture the lowest order backreaction effects.

\section{Backreaction terms} \label{backterms}

Here we want to consider the backreaction effects of the inhomogeneities onto the homogeneous variables. To this end we have to find the complete observables associated to a homogeneous variable up to second order. (The first order of such a complete observable vanishes.)

Omitting in (\ref{nov221}) with $f$ a function of zeroth order all terms of third order and higher we find 
\ba\label{jan181} 
{}^{[2]}F_{[f;T^K]}(\tau) &\simeq& {}^{(0)}F_{[f;T^K]}(\tau) + G + I+ J
\ea
where
\ba\label{jan182}
{}^{(0)}F_{[f;T^K]}(\tau) &=& \sum_{q=0}^\infty \frac{1}{q!} \{ f, \, {}^{(0)}\tilde C_0(0)\}_{q}    (\tau-T^0(0))^q =:\alpha_{free}^{(\tau-T^0(0))}(f)      
\q\q\q \text{and}
\nn 
\\
G &=&
\sum_{\hat k_1} \{\, \alpha_{free}^{(\tau-T^0(0))}(f) , {}^{(1)}\tilde C_{\hat K_1}(\hat k_1)\}\,\, (-T^{\hat K_1}(\hat k_1)) 
\,\,+ \nn
\\ 
&&\sum_{\hat k_1, \hat k_2} \frac{1}{2!}\,{}^{(0)}\{ \{\, \alpha_{free}^{(\tau-T^0(0))}(f) , {}^{(1)}\tilde C_{\hat K_1}(\hat k_1)\}, {}^{(1)}\tilde C_{\hat K_2}(\hat k_2)\}\,\, (-T^{\hat K_1}(\hat k_1))  (-T^{\hat K_2}(\hat k_2)) \q . \q \q\q 
\ea
Hence $G$ is the gauge invariant extension to second order with respect to the $\tilde C_{\hat K}(\hat k)$ constraints of the zeroth order term. The last two terms in (\ref{jan181}) are given by
\ba\label{jan183}
I&=&  \sum_{q=0}^\infty \,\, \sum_{q_1+q_2+1=q} {}^{(2)}\{\,\{\{f,\,{}^{(0)}\tilde C_{0}(0)\}_{q_1},\, {}^{(2)}\tilde C_{0}\},\, {}^{[2]}\tilde C_{0}(0) \}_{q_2}\, (\tau-T^0(0))^q \nn\\
J&=& \sum_{\hat k_1}\,   {}^{(1)} \{ I,\, {}^{(1)}\tilde C_{\hat K_1}(\hat k_1) \}\,\,  (-T^{\hat K_1}(\hat k_1)) \,\,+\nn
\\
 &&
\sum_{\hat k_1, \hat k_2}\, {}^{(0)}\{\,  {}^{(1)}\{I,\,{}^{(1)}\tilde C_{\hat K_1}(\hat k_1)\},\, \,{}^{(1)}\tilde C_{\hat K_2}(\hat k_2)\} \,\, (-T^{\hat K_1}(\hat k_1))  (-T^{\hat K_2}(\hat k_2))  \q .\q\q
\ea
Using the identity
\ba\label{jan184}
\frac{t^{q_1+q_2+1}}{(q_1+q_2+1)!}=\frac{1}{q_1!}\frac{1}{q_2!}\int_0^t \bd s \, (t-s)^{q_1} s^{q_2} 
\ea
we can rewrite the term $I$ as
\ba\label{jan185}
I=\int_0^{\tau-T^0(0)}\bd s  \,\,\, \alpha_{free}^{(\tau-T^0(0)-s)}( \,\{ \alpha_{free}^s(f),\,{}^{(2)}\tilde C_0(0)\}\,)
\ea
where we define the free evolution of a higher order term ${}^{(m)}g$ as
\ba\label{jan186} 
\alpha_{free}^t({}^{(m)}g)=\sum_{q=0}^\infty \frac{1}{q!} \,\,{}^{(m)}\{\,{}^{(m)}g, \,{}^{[2]}\tilde C_0(0)\}_q\,\,t^q \q .
\ea
We have also for the free evolution the factorization property $\alpha^t_{free}(f\cdot g)=\alpha^t_{free}(f)\cdot\alpha_{free}^t(g)$, hence the free evolution of a higher order term is determined by the free evolution of its zero and first order constituents. Note that for the free evolution of a first order term one can drop all second and higher order terms which might arise, i.e.  
\ba\label{jan187}
{}^{(1)}\{{}^{(0)}f \,\,{}^{(1)}g,\,{}^{(0)}\tilde C_{0}(0)
+
{}^{(2)}\tilde C_{0}(0)\}
=
\{{}^{(0)}f,\, {}^{(0)}\tilde C_0(0)\}\,{}^{(1)}g
+
\,{}^{(0)}f\{{}^{(1)}g,\, {}^{(2)}\tilde C_0(0)\}  \q .
\ea
In particular the free propagation for the linear terms is linear, i.e. we can express the free evolution of a first order term via the propagator functions used in section \ref{scalarmodes}.

By using that 
\ba\label{jan191}
\{{}^{[2]}\tilde C_0(0), T^{\hat K}(\hat K)\}=O(C)+O(2)\q\q\text{and} \q\q \{{}^{[2]}\tilde C_0(0), \,{}^{(1)}\tilde C_{\hat K}(\hat k)\}=O(C^2)+O(2)
\ea
we obtain for the term $I+J$
\ba\label{jan192}
I+J &\simeq & \int_0^{\tau-T^0(0)}\bd s  \,\,\, \alpha_{free}^{(\tau-T^0(0)-s)}\bigg( \,\{ \alpha_{free}^s(f),\,{}^{(2)}\tilde C_0(0)\} \,+ \nn\\
&&
\quad\quad\quad\quad\quad
{}^{(1)} \big{\{}  
\{ \alpha_{free}^s(f),\,{}^{(2)}\tilde C_0(0)\}
, {}^{(1)}\tilde C_{\hat K_1}(\hat k_1)  
\big{\}} (-T^{\hat K_1}(\hat k_1))
\,  \, + \nn \\
&&
\quad\quad\quad\quad\quad
\frac{1}{2!} {}^{(0)}\big{\{}
\big{\{}  
\{ \alpha_{free}^s(f),\,{}^{(2)}\tilde C_0(0)\}
, {}^{(1)}\tilde C_{\hat K_1}(\hat k_1)  
\big{\}}
, {}^{(1)}\tilde C_{\hat K_2}(\hat k_2)
 \big{\}} 
 (-T^{\hat K_1}(\hat k_1)) (-T^{\hat K_2}(\hat k_2))
\,\bigg). \nn\\
\ea
The second and third term project out of $\{ \alpha_{free}^s(f),\,{}^{(2)}\tilde C_0(0)\}$ all terms proportional to the clock variables $T^{\hat K}(\hat k)$. Furthermore because of (\ref{jan191}) one does not need to evolve terms proportional to the linearized constraints ${}^{(1)}\tilde C_{\hat K}(\hat k)$, hence one is left with the evolution of the scalar mode and the two tensor modes.

In summary we learn that the gauge invariant second order backreaction effect ${}^{[2]}F_{[f;T^K]}(\tau)$ consists of two pieces: one is the gauge invariant extension of the homogeneous term ${}^{(0)}F_{[f;T^K]}(\tau)$ to the appropriate order, the other piece comes about through the ``interaction'' of the homogeneous variables with the second order part  of the time generating constraint ${}^{(2)}\tilde C_0(0)$. Here one needs to consider only the first order gauge invariant terms (i.e. the first order physical modes) that arise in this interaction.

Note that for the second order complete observable ${}^{[2]}F_{[f;T^K]}(\tau)$ associated to a zeroth order function $f$ we do not need to consider the integrated vector constraints (related to the linearization instabilities). Since these start at second order and ${}^{[2]}F_{[f;T^K]}(\tau)$ does not contain a first order term, this second order complete observables is already invariant modulo second order terms with respect to the integrated vector constraints.

Higher order complete observables can be calculated by expanding the power series for complete observables (\ref{nov221}) systematically and by using the identity (\ref{jan184}) repeatedly. This will result in a Feynman--graph like expansion, that is a sum of terms describing different interaction processes generated by the higher order terms of the constraint $\tilde C_0(0)$.

In the next section  we will calculate for a very simple model the lowest order backreaction effect onto the isotropic and homogeneous geometry explicitly.

\section{ Bianchi I as a model for perturbations} \label{bianchi}

Here, similar to \cite{hector} we will consider anisotropic but homogeneous cosmologies of Bianchi--I--type as a perturbation of isotropic and homogeneous cosmologies. We will calculate the lowest order effect of the anisotropies onto the isotropic and homogeneous variables. The Bianchi I model (with a massless scalar field) is solvable, therefore one can compare the results to the exact model. 

\subsection{The model}
For the Bianchi I model the connection $A_{a}^{\; j}$ is given by $A_{a}^{\; j}= \beta \diag(A_1, A_2, A_3)$ and the canonical momentum $E^{a}_{\;j}$ is given by $E^{a}_{\;j}=\beta^{-1} \diag(E_1, E_2, E_3)$ (see \cite{hector}). Further on we assume that there is a massless and homogeneous scalar field  $\Phi$ and its canonical momentum $\Pi$.\\
The Gauss constraint and the diffeomorphism constraint are trivially fulfilled, the Hamiltonian constraints reduces to the following (weight 2 version)
\begin{eqnarray}\label{bianchicons}
C  =  \frac{2 \beta^2}{\kappa} (E_1 A_1 E_2 A_2 + E_1 A_1 E_3 A_3 + E_2 A_2 E_3 A_3) + \frac{1}{2 \gamma} \Pi^2
\end{eqnarray}
where $\beta=i/2$ is the Immirzi parameter and $\kappa$ and $\gamma$ are coupling constants.
The symplectic structure is given by:
\begin{eqnarray}
\{A_i, E_j  \}  =  \kappa\delta_{ij}  \q\q\q\q\q\q\q\q
\{\Phi, \Pi  \}  =  \gamma   \q   
\end{eqnarray}
where $i,j=1,2,3$.

\subsection{Expansion around the isotropic sector}
In order to expand the model around the isotropic sector we define
\begin{eqnarray}
 A:=\frac{1}{3}\sum\limits_{i}A_{i} & & E:=\frac{1}{3}\sum\limits_{i}E_{i}\\
  a_{i} :=A_i-\frac{1}{3}\sum\limits_{k}A_{k} & & e_{i} :=E_i - \frac{1}{3}\sum\limits_{k}E_{k}  \q .
\end{eqnarray}
This can be implemented through a projector $\cp$ which projects a phase space function $f$ onto its averaged value: $\cp(f)(A_1,A_2,A_3;E_1,E_2,E_3)=f(A,A,A;E,E,E)$. We will call the isotropic variables A and E zeroth order variables and the ``fluctuations" $a_{i}$ and $e_{i}$ first order variables.\\
The symplectic structure in these variables reduces to
\begin{eqnarray}
\{A, E  \} = \frac{1}{3}\kappa, & & \{a_{i}, e_{j}\} = (\delta_{ij}-\frac{1}{3})\kappa, \qquad \{\Phi, \Pi \} = \gamma  \q .
\end{eqnarray}
The fluctuation variables are not completely independent of each other and fulfil the following relations (by construction):
\begin{eqnarray}
\sum\limits_{i}a_{i}= 0 & & \sum\limits_{i}e_{i} = 0  \q .
\end{eqnarray}
In these variables the Hamiltonian constraint can be split into 4 parts:
\begin{eqnarray}
C & = & \;^{(0)}C + \;^{(2)}C + \;^{(3)}C + \;^{(4)}C\\
\;^{(0)}C & = & \frac{6 \beta^2}{\kappa}A^2 E^2 + \frac{1}{2\gamma}\Pi^2\\
\;^{(2)}C & = & \frac{2 \beta^2}{\kappa} A E \sum\limits_{i}a_{i}e_{i}-\frac{\beta^2}{\kappa}A^2\sum\limits_{i}e_{i}e_{i}-\frac{\beta^2}{\kappa}E^2\sum\limits_{i}a_{i}a_{i}\\
\;^{(3)}C & = & - \frac{2 \beta^2}{\kappa}(A\sum\limits_{i}a_{i}e_{i}e_{i} + E \sum\limits_{i}a_{i}a_{i}e_{i})\\
\;^{(4)}C & = & \frac{\beta^2}{\kappa}\sum\limits_{i\neq j}a_{i}e_{i}a_{j}e_{j}  \q .
\end{eqnarray}

\subsection{Exact solution}
Fortunately this model can be solved exactly and we can compare the exact solution with the one that we will derive perturbatively later. We choose lapse $N=1$ and can obtain the exact solution by solving the following first order system of differential equations.
\begin{eqnarray}
\dot{A}_{i} & := &\{ A , C \} = 2 \beta^2 A_{i}(A_j E_j + A_k E_k)\\
\dot{E}_{i} & := &\{ E , C \} = -2 \beta^2 E_{i}(A_j E_j + A_k E_k)\\
\dot{\Phi} & := &\{ \Phi , C \} = \Pi\\
\dot{\Pi} & := & \ \{\Pi, C  \} = 0
\end{eqnarray}
where the indices $i,j,k$ on the right hand of these equation are mutually different. (The Einstein sum convention does not apply here and throughout this section.)
The dot refers to derivative in coordinate time $t$ (with the choice $N=1$).
One can see that $A_i E_i$ is a Dirac-observable, i.e. Poisson-commutes with C for $i=1,2,3$.\\
We can easily solve this system of differential equations and obtain the following solutions:
\begin{eqnarray}
A_i(t) & = & A_{i}\exp[-2\beta^2 A_i E_i t] \exp[2\beta^2\sum\limits_{j}A_j E_j t] \label{ex1} \\
E_{i}(t) & = & E_{i}\exp[2\beta^2A_i E_i t] \exp[-2\beta^2\sum\limits_{j}A_j E_j t]\label{ex2} \\
\Phi(t) & = & \Pi t + \Phi\label{ex3} \\
\Pi(t) & = & \Pi \label{ex4}
\end{eqnarray}
To calculate complete observables we have to specify a clock variable in order to get rid of the physically meaningless coordinate time t. For our purposes it is convenient to choose $T = \frac{\Phi}{\Pi}$ because it evolves linearly in coordinate time: $T(t) = t + \frac{\Phi}{\Pi}$. Inverting this relation and inserting it into (\ref{ex1}), (\ref{ex2}), (\ref{ex3}), (\ref{ex4}) leads to the following complete observables:
\begin{eqnarray}
F_{[A_i, T=\frac{\Phi}{\Pi}]}(\tau) & = & A_{i}\exp[-2\beta^2 A_i E_i (\tau - \frac{\Phi}{\Pi})] \exp[2\beta^2\sum\limits_{j}A_j E_j (\tau - \frac{\Phi}{\Pi})] \label{obs1} \\
F_{[E_i, T=\frac{\Phi}{\Pi}]}(\tau) & = & E_{i}\exp[2\beta^2A_i E_i (\tau - \frac{\Phi}{\Pi})] \exp[-2\beta^2\sum\limits_{j}A_j E_j (\tau - \frac{\Phi}{\Pi})]\label{obs2} \\
F_{[\Phi, T=\frac{\Phi}{\Pi}]}(\tau) & = & \Pi \tau \label{obs3} \\
F_{[\Pi, T=\frac{\Phi}{\Pi}]}(\tau) & = & \Pi \label{obs4} \q .
\end{eqnarray}
For the isotropic variables the observables read as follows:
\begin{eqnarray}
F_{[A, T=\frac{\Phi}{\Pi}]}(\tau) & = & \frac{1}{3}\sum\limits_{i}A_{i}\exp[-2\beta^2A_{i}E_{i}(\tau-\frac{\Phi}{\Pi})]\exp[2\beta^2\sum\limits_{j}A_{j}E_{j}(\tau-\frac{\Phi}{\Pi})] \label{obsc}\\
F_{[E, T=\frac{\Phi}{\Pi}]}(\tau) & = & \frac{1}{3}\sum\limits_{i}E_{i}\exp[2\beta^2A_{i}E_{i}(\tau-\frac{\Phi}{\Pi})]\exp[-2\beta^2\sum\limits_{j}A_{j}E_{j}(\tau-\frac{\Phi}{\Pi})] \label{obsp}         \q .
\end{eqnarray}
To be able to compare these exact results with the perturbative calculation later we can Taylor--expand (\ref{obsp}) around $E_{i} = E, A_{i} = A$. For the complete observable associated to E we obtain:
\begin{eqnarray}
F_{[E, T=\frac{\Phi}{\Pi}]}(\tau) & = &  \exp[-\omega(\tau-\frac{\Phi}{\Pi})]\times \label{Taylor2} \nonumber\\
& & \times \Big[ E + \frac{2}{3} \beta^2 A (\tau -\frac{\Phi}{\Pi})\sum\limits_{i}e_i e_i - \frac{2}{3} \beta^2 E  (\tau -\frac{\Phi}{\Pi})\sum\limits_{i} a_i e_i  \,\, + \nonumber \\
& & \quad    \frac{1}{6}\beta^2 \omega A (\tau - \frac{\Phi}{\Pi})^2 \sum\limits_{i}e_i e_i  + \frac{1}{6} \beta^2 \omega \frac{E^2}{A} (\tau - \frac{\Phi}{\Pi})^2  \sum\limits_{i} a_i a_i   
 +  \frac{1}{3} \beta^2 \omega E (\tau - \frac{\Phi}{\Pi})^2  \sum\limits_{i} a_i e_i  \Big] \nonumber \\
& & + O(3) \q .
\end{eqnarray}
where we introduced the abbreviation $\omega = 4\beta^2 A E$ and $O(3)$ denotes terms of order 3 and higher in the anisotropic fluctuations. The first order term vanishes due to the condition $\sum\limits_{i} e_i = \sum\limits_{i} a_i =0$.


\subsection{Using the perturbative approach}
Now that we know the exact solution, we can try to reproduce this results order per order using the perturbative approach to complete observables. Here we will only consider the lowest non--trivial order of the complete observable associated to the homogeneous variable $E$. 
In the first step we have to specify a clock function. If we choose $T=\frac{\Phi}{\Pi}$ (in order to fulfil $\{T, C  \} = 1$) as a clock, the analysis gets as simple as possible, because $\tilde{C}:= (\{T, C  \})^{-1}C= C$.\\
In this case the complete observable associated to an arbitrary phase space function $f$ is (formally) given by
\begin{eqnarray}
F_{[f, T=\frac{\Phi}{\Pi}]}(\tau) & = & \sum\limits_{0}^{\infty}\frac{(\tau - \frac{\Phi}{\Pi})^{n}}{n!}\{f, C  \}_{n} \label{complete} \nonumber\\
& = & \sum\limits_{0}^{\infty}\frac{(\tau - \frac{\Phi}{\Pi})^{n}}{n!}\{f, \;^{(0)}C + \;^{(2)}C + \;^{(3)}C + \;^{(4)}C  \}_{n}  \q .
\end{eqnarray} 
We can evaluate this sum order per order in the fluctuation variables. For all zeroth order quantities $f$ we obtain
\begin{eqnarray}
\;^{(0)}F_{[f,T=\frac{\Phi}{\Pi}]}(\tau) & = & \sum\limits_{n=0}^{\infty}\frac{(\tau-\frac{\Phi}{\Pi})^n}{n!}\{f, \;^{(0)}C  \}_{n} = :\flow_{\;^{(0)}C}^{t}(f)\\
\;^{(1)}F_{[f,T=\frac{\Phi}{\Pi}]}(\tau) & = & 0\\
\;^{(2)}F_{[f,T=\frac{\Phi}{\Pi}]}(\tau) & = & \sum\limits_{n=1}^{\infty}\sum\limits_{n_1+n_2=n-1}\frac{(\tau-\frac{\Phi}{\Pi})^n}{n!}\;^{(2)}\{ \{ \{f, \;^{(0)}C  \}_{n_1}, \;^{(2)}C \}, \;^{(0+2)}C \}_{n_2} \label{approx02} \q .
\end{eqnarray}
For first order quantities $f$ we get
\begin{eqnarray}
\;^{(1)}F_{[f,T=\frac{\Phi}{\Pi}]}(\tau) & = & \sum\limits_{n=0}^{\infty}\frac{(\tau-\frac{\Phi}{\Pi})^n}{n!}\:^{(1)}\{f, \;^{(0+2)}C  \}_{n}=:\flow_{free}^{\tau-\frac{\Phi}{\Pi}  } (f)\label{approx11}
\end{eqnarray}
Expression (\ref{approx02}) can be interpreted as follows: A zeroth order quantity $f$ propagates with respect to the ``free'' Hamiltonian constraint $\;^{(0)}C$, then there is an ``interaction'' with the second order Hamiltonian constraint $\;^{(2)}$C which yields terms quadratic in the fluctuations. After the interaction zeroth order quantities evolve according to the ``free'' Hamiltonian constraint $\;^{(0)}C$ and first order quantities according to the sum of the zeroth and second order Hamiltonian constraint ${}^{(0)}C+{}^{(2)}C$ (where one can ignore all higher than first order quantities that appear in evolving the first order quantity).\\
The interpretation of (\ref{approx11}), which can be seen as the ``free'' propagation of a first order quantity $f$, is similar.\\
We will compute the second order of $F_{[E, T=\frac{\Phi}{\Pi}]}(\tau)$. Using identity (\ref{jan184}) the expression (\ref{approx02}) can be reformulated as follows:
\begin{eqnarray}
\;^{(2)}F_{[f,T=\frac{\Phi}{\Pi}]}(\tau) & = & \int\limits_{0}^{\tau-\frac{\Phi}{\Pi}}\flow_{free}^{\tau-\frac{\Phi}{\Pi}-t}\Big[\Big\{\flow_{\;^{(0)}C}^{t}(E), \;^{(2)}C   \Big\}   \Big] \label{int} \q .
\end{eqnarray}
The first step is to calculate
\begin{eqnarray}
\flow_{\;^{(0)}C}^{t}(A), & \qquad & \flow_{\;^{(0)}C}^{t}(E) , \label{iso}\\
\; \nonumber \\
\flow_{free}^{t}(a_i),  & \qquad &  \flow_{free}^{t}(e_i).  \label{aniso}
\end{eqnarray}
The first two quantities, (\ref{iso}), are the solutions to the differential equations:
\begin{eqnarray}
\dot{A} & : = & \{ A,\;^{(0)}C \} =  4\beta^2 A^2 E  \label{DE1}\\
\dot{E} & : = & \{ E,\;^{(0)}C \} =  -4\beta^2 A E^2  \label{DE2}  \q .
\end{eqnarray}
These can easily be found to be
\begin{eqnarray}
\flow_{\;^{(0)}C}^{t}(A) & = & A\exp(\omega t) \label{A0}\\
\flow_{\;^{(0)}C}^{t}(E) & = & E\exp(-\omega t)\label{E0}\\
\omega & := & 4 \beta^2 A E \label{omega}  \q\q\q\q  .
\end{eqnarray}
The second set of quantities can be found by solving
\begin{eqnarray}
\dot{a}_{k} & = & \{a_k, {}^{(2)}C\} = 2\beta^2 A E  a_{k} - 2\beta^2 A^2 e_{k}\\
\dot{e}_{k} & = & \{e_k, {}^{(2)}C\} = -2\beta^2 A E  e_{k} + 2 \beta^2 E^2 a_{k}
\end{eqnarray}
where the time dependence of the homogeneous variables is given by $A(t)=\flow_{\;^{(0)}C}^{t}(A)$ and $E(t)=\flow_{\;^{(0)}C}^{t}(E)$. The solution to this set of differential equations is given by
\begin{eqnarray}
\flow_{free}^{t}(a_k) & = & a_{k}\exp[\omega t] - \frac{\omega}{2}(a_{k}+\frac{A}{E}e_{k})t\exp[\omega t] \label{sol1} \\
\flow_{free}^{t}(e_k) & = & e_{k}\exp[-\omega t] + \frac{\omega}{2}(e_{k} + \frac{E}{A}a_{k})t\exp[-\omega t] \label{sol2}  \q .
\end{eqnarray}
Now we can calculate (\ref{int}) step by step:
\begin{eqnarray}
\Big\{ \flow_{\;^{(0)}C}^{t}(E), \;^{(2)}C  \Big\} & = & \frac{1}{3}\exp[-\omega t] \Big[ 2\beta^2 A (1-\omega t)\sum\limits_{i}\!\! e_i e_i  -2 \beta^2 E \sum\limits_{i}\!\! e_{i} a_{i} 
 + \quad 2 \beta^2 \omega t \frac{E^2}{A}\sum\limits_{i}  \!\! a_i a_i   \Big] \nonumber \\
\; \nonumber \\
\flow_{free}^{\tau-\frac{\Phi}{\Pi}-t}\Big[\Big\{\flow_{\;^{(0)}C}^{t}(E), \;^{(2)}C   \Big\}   \Big]
 & = & 
\frac{2}{3}\beta^2 A \exp[-\omega(\tau-\frac{\Phi}{\Pi})]\times\nonumber\\
& & \times \Big[ \sum\limits_{i} e_{i} e_{i} -\frac{E}{A} \sum\limits_{i} a_{i} e_{i}
 + \nonumber \\
& & \qquad 
 \omega (\tau -\frac{\Phi}{\Pi})\big[ \frac{3}{2}\sum\limits_{i} e_{i} e_{i} 
+ \frac{E}{A} \sum\limits_{i}e_{i}a_{i} 
- \frac{1}{2} \frac{E^2}{A^2} \sum\limits_{i} a_{i} a_{i}  \big]  \, + \nonumber \\
& & \qquad 
 \frac{1}{2}\omega^2(\tau-\frac{\Phi}{\Pi})^2 \big[ \sum\limits_{i} e_{i} e_{i} 
+ \frac{E^2}{A^2} \sum\limits_{i} a_{i} a_{i} 
+ 2 \frac{E}{A} \sum\limits_{i} e_{i} a_{i} \big]\,+ \nonumber \\
& & \quad  \quad  \omega t \big[ -\frac{5}{2}\sum\limits_{i} e_{i} e_{i} -\frac{E}{A}\sum\limits_{i}e_{i}a_{i} +\frac{3}{2}\frac{E^2}{A^2}\sum\limits_{i} a_{i} a_{i}+ \nonumber \\
& & \qquad \qquad 
-2 \omega(\tau - \frac{\Phi}{\Pi})\big[ \sum\limits_{i} e_{i} e_{i} + \frac{E^2}{A^2}\sum\limits_{i} a_{i} a_{i} + 2 \frac{E}{A} \sum\limits_{i} e_{i} a_{i}  \big]   \big]\,+\nonumber \\
& & \quad \quad  \frac{3}{2}\omega^2 t^2 \big[ \sum\limits_{i} e_{i} e_{i} + \frac{E^2}{A^2}\sum\limits_{i} a_{i} a_{i} + 2 \frac{E}{A} \sum\limits_{i} e_{i} a_{i} \big]   \Big]  \q .
\end{eqnarray}

\begin{figure} 
   \centering
	\psfrag{ylabel}{$E/[m^2]$}
	\psfrag{xlabel}{$\tau-\frac{\Phi}{\Pi}/[m^{-2}]$}
       \includegraphics[scale=0.9]{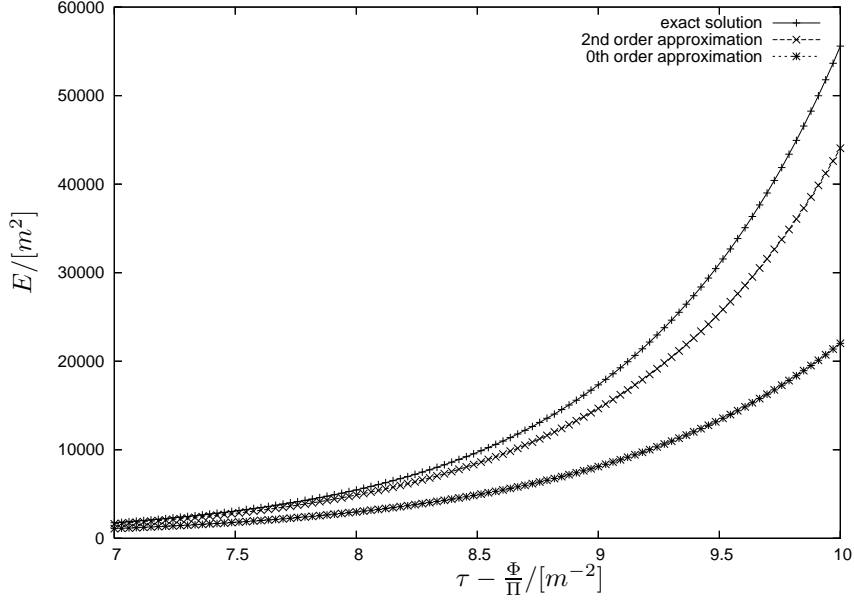}
    \caption{\label{EinBild} Bianchi I as  a model for perturbations: Here we compare the exact complete observable evaluated on a phase space point $(A_i,E_i,\Phi,\Pi)$ to the zeroth and second order approximation. The phase space point is given by $A=1,\, E=1m^2,\,  a_1=a_2=0.1, \, a_3=-0.2,\, e_1=e_2=0.1 m^2, e_3=-0.2 m^2$ (where we set the speed of light to $c=1$ and the coordinates do not carry units). The momentum $\Pi$ is determined by the constraint (\ref{bianchicons}). Choosing a value for $\Phi$ defines the time $\tau$ for which the above values are taken as initial values. }
  \label{figure1}
\end{figure}

This gives the second order contribution 
\begin{eqnarray}\label{perturb2}
\;^{(2)}F_{[E,T=\frac{\Phi}{\Pi}]}(\tau) 
& = & 
\int\limits_{0}^{\tau-\frac{\Phi}{\Pi}}dt
\flow_{free}^{\tau-\frac{\Phi}{\Pi}-t}\Big[\Big\{\flow_{\;^{(0)}C}^{t}(E), \;^{(2)}C   \Big\}   \Big] \nonumber \\
& = & 
\!\! \exp[-\omega(\tau-\frac{\Phi}{\Pi})]\times  
\Big[\frac{2}{3} \beta^2 A (\tau -\frac{\Phi}{\Pi})\sum\limits_{i}\!\! e_i e_i - \frac{2}{3} \beta^2 E  (\tau -\frac{\Phi}{\Pi})\sum\limits_{i}\!\!  a_i e_i  
 \nonumber \\
& & \quad \!\!
   +   \frac{\beta^2}{6} \omega A (\tau - \frac{\Phi}{\Pi})^2 \sum\limits_{i}\!\! e_i e_i  +  
\frac{\beta^2}{6} \omega \frac{E^2}{A} (\tau - \frac{\Phi}{\Pi})^2  \sum\limits_{i}\!\! a_i a_i 
+  \frac{\beta^2}{3} \omega E (\tau - \frac{\Phi}{\Pi})^2 \sum\limits_{i}\!\! a_i e_i  \Big]. \nn\\
\end{eqnarray}
As expected, (\ref{perturb2}) coincides with (\ref{Taylor2}), the result we obtained by just Taylor--expanding the exact expression. This simple model shows that it is, at least in principle, possible to calculate backreaction effects in cosmological applications of general relativity using the methods developed in this work.


As an illustration we compare in figure \ref{EinBild} the exact complete observable with the zeroth and the second order complete observable associated to $E$. As one can see there the (relative) difference of the second order approximation to the zeroth order approximation grows with ``time'' $\tau$, as does the difference of the second order approximation to the exact solution.



\section{Discussion and Conclusions}\label{discuss}

We presented a gauge invariant canonical scheme for perturbations around symmetry reduced sectors of gauge systems. This scheme is applied to general relativiy, in particular to perturbations around the cosmological sector. It can be used to calculate the dynamics up to arbitrary high order in the fluctuation  variables. 

The central objects in this perturbative scheme are complete observables. These complete observables are gauge invariant (i.e. Dirac) observables. In the canonical formalism this means that the complete observables have to be invariant under (coordinate) time reparametrizations, i.e. constants of motions. Nevertheless one can express the dynamics of the theory using the complete observables. To this end one has to choose a set of physical clocks. Evolution of dynamical entitities can then be understood as an evolution in relation to the clocks.  One can even define generalized Hamiltonians, that is gauge invariant phase space functions that generate this evolution for the complete observables.

Different choices of clocks can be interpreted as different setups for a physical measurement.  We gave explicit formulas relating complete observables associated to different choices of clocks in section \ref{trafo}.
In particular the clocks define the hypersurfaces over which the averaging (\ref{proj}) is defined. The complete observables evaluated on a phase space point (i.e. evaluated on a spacetime satisfying the Einstein equations) give the values of the partial observables, such as the averaged densitized triad $E$ and its fluctuations $e^a_b$, on the hypersurface determined by $\{T^K(k)=\tau^K(k)\}$. Gauge invariance of the complete observables ensures that the value of the complete observable does not change if we evaluate the complete observable on spacetimes related by diffeomorphisms.   From the complete observables associated to all the phase space variables (which will give a overcomplete basis of observables) one can calculate the values of all kinds of physical observables on the hypersurface $\{T^K(k)=\tau^K(k)\}$. For instance one might wish to consider instead of the averaged densitized triad some  averaged function of the (spacetime\footnote{The four metric can be computed by using lapse and shift functions defined in section \ref{lapseandshift}, see also \cite{bd2}.}) metric components: In this case one has first to express the partial observable in question as a function $f$ of the zeroth and first order phase space variables and then to consider the complete observable associated to this function (expanded to the appropriate order). Note that this complete observable can be expressed as the same function $f$ of the complete observables associated to the zeroth and first order complete observables. 

We considered in more detail first and second order complete observables. Higher complete observables can be calculated by similar methods. The zeroth order complete observables coincide with the Dirac observables of the symmetry reduced model. The dynamics of the first order complete observables is related to the theory of cosmological linear perturbations (see \cite{MFB} and references therein), the difference is that in our case dynamics is expressed with respect to a (global) clock variable, which can however be related to choosing a (zeroth order) lapse function. 

Second order complete observables associated to zeroth order partial observables describe backreaction effects. We calculated these backreaction effects in section \ref{bianchi} for a simple model. To allow for an explicit expression for these effects in interesting (inflationary) scenarios further approximations, as for instance slow roll or long wave length approximations, need to be implemented. This can be done by introducing small parameters, which characterize these approximations \cite{jt}. The power series for the complete observables has then to be expanded to a certain order in the fluctuation variables and to a certain order in these new parameters.

A perturbative scheme around a whole phase space sector has several advantages compared to an expansion around a fixed phase space point. Firstly, some of the degrees of freedom, namely these corresponding to the symmetry reduced sector are treated non--perturbatively. These degrees of freedom are used to define the clock in relation to which we express the dynamics of the theory. Compared to defining a field theory on a fixed background, where the background is used to define time, the clock in this approach is a fully dynamical entity, as one would expect in the full theory. Indeed one would expect problems with gauge consistency to higher order  if one uses a background time, 
since such a time cannot take into account backreaction effects onto the physical clock. 

Also a quantization of the theory should not only quantize the perturbations on a fixed background but rather consider the quantized perturbations on a quantized geometry, described by the (quantized) symmetry reduced sector.  The variables describing the symmetry reduced sector contribute for instance to the Hamiltonian generating the physical time evolution. From this perspective it might be fruitful to reconsider certain concepts, like energy and vacua, from quantum field theory on curved spacetime.


As in the case for perturbations around a fixed background, where the background is given in certain coordinates, the symmetry reduced sector is described by using a certain type of coordinates. That is, there exist configurations which are not included in the symmetry reduced sector, but which are nevertheless physically symmetric. These configurations can be obtained by a ``non--symmetric'' coordinate transformation from a symmetric configuration. Degrees of freedom gained in this manner correspond to gauge degrees of freedom. 

However by expanding around a sector describing a family of solutions in different coordinate systems one can investigate the non--perturbative effects of such gauge degrees of freedom compared to an expansion around a smaller sector (which could be just one phase space point), which would treat some of these gauge degrees of freedom perturbatively \cite{wip}. That is we can embed symmetry reduced models into each other and explore in this way the reliability of symmetry reduction.

\vspace{1cm}

\begin{appendix}

~\\
{\huge{  \bf Appendix }}

\section{Tensor mode decomposition} \label{tensor modes}

Similar to the longitudinal and transversal modes for a vector field on $\Rl^3$ one can introduce tensor modes for a tensor field. For a proof of the completeness of these modes, see \cite{mcp4}. To begin with we define the projector onto the transversal modes of a vector field by
\be
(p\cdot v)_a:=p_a^b\cdot v_b:=(\delta_a^b+W^{-2}\cdot \partial_a \partial_b)\cdot v_b \q .
\ee
This allows us to introduce the following basis of tensor modes:
\begin{xalignat}{2}\label{lingravBasis2}
&({}^{LT}P\cdot T)_{ab}=(\delta_a^c-p_a^c)\cdot p_b^d \cdot T_{cd} && \mbox{2 left long. right transv. modes}\nn \\
&({}^{LL}P\cdot T)_{ab}=(\delta_a^c-p_a^c)\cdot(\delta_b^d-p_b^d) \cdot T_{cd} && \mbox{1 left and right long. mode}\nn \\
&({}^{TL}P\cdot T)_{ab}=p_a^c\cdot (\delta_b^d-p_b^d) \cdot T_{cd} &&\mbox{2 left transv. right long. modes}\nn \\
&({}^{T}P\cdot T)_{ab}= \tfrac{1}{2}p_{ab}\cdot p^{cd}\cdot T_{cd} && \mbox{1 symm. transv. trace part mode} \nn \\
&({}^{AT}P\cdot T)_{ab}=\tfrac{1}{2}(p_a^c\cdot p_b^d-p_b^c\cdot p_a^d)\cdot T_{cd} &&\mbox{1 antisymm. transv. mode} \nn \\
&({}^{STT}P\cdot T)_{ab}= \tfrac{1}{2}(p_a^c\cdot p_b^d +p_b^c\cdot p_a^d-p_{ab}p^{cd})\cdot T_{cd} &&\mbox{2 symm. transv. tracefree modes}
\end{xalignat}
 Using the projector property $p\cdot p=p$, it is easy to see that the 
 projectors ${}^{X}P$ are orthogonal to each other and satisfy
 ${}^{X}P\cdot{}^{Y} P=\delta^{XY}\,\,{}^{X}P$. Furthermore the set of projectors is complete, that is
\ba
\sum_X {}^{X}P_{ab}^{cd}=\delta^c_a \delta^d_b  \q .
\ea
We will denote the tensor modes by ${}^{X}T_{ab}:=({}^{X}P\cdot T)_{ab}$.

\section{First order perturbations: scalar modes in longitudinal gauge}\label{longo}

Here we will derive the equations of motions for the first order scalar perturbations with the choice (\ref{choiceLong}) 
\ba\label{choceLonga}
T^{Ga} &=&\epsilon^{abc}e_{bc}  \nn\\
          &=&\epsilon^{abc}({}^{AT}e_{bc}+{}^{LT}e_{bc}+{}^{TL}e_{bc}) \nn\\
T^{Da} &=& -W^{-2} ( -\tfrac{1}{2} W^{-2}\partial^a\partial_d\partial_e +\tfrac{1}{2} \partial^a \delta_{de} -\partial_e \delta^a_d -\partial_d \delta^a_e)e^{de} \nn\\
&=& 
-W^{-2} ( -\partial^a\,{}^{LL}{e^d}_d+\frac{1}{2}\partial^a\,{}^{T}{e^d}_d -\partial_e\,{}^{TL}e^{ae}-\partial_d\,{}^{LT}e^{da})  \nn\\
T^0 &=&       W^{-2}(\tfrac{1}{2} \delta^{cd} +\tfrac{3}{2}W^{-2}\partial^c\partial^d)a_{cd}                                                        \nn\\
&=&W^{-2}(\tfrac{1}{2}\,{}^{T}{a^d}_d-\,{}^{LL}{a^d}_d)  \q .
\ea 
for the clock variables. As we will see this choice is related to the longitudinal gauge: In the notation of section \ref{scalarmodes} we have
\ba
[[T^{Da}]]=[[  -W^{-2} ( -\tfrac{1}{2} W^{-2}\partial^a\partial_d\partial_e +\tfrac{1}{2} \partial^a \delta_{de} -\partial_e \delta^a_d -\partial_d \delta^a_e)e^{de} ]] \simeq 0 \q .
\ea 
(Note that the brackets $[[\, \cdot \,]]$ now refer to complete observables with respect to the clock variables (\ref{choceLonga}).) We want to translate this condition on the triad perturbation to a condition on the spatial metric $q_{ab}$. With the relation 
\ba\label{jan201}
\text{det}(q)q^{ab}=\beta^2 E^a_j E^b_j
\ea
between the densitized triad variables $E^a_j$ and the inverse metric $q^{ab}$ (where $\text{det}(q)$ is the determinant of the metric) we get
\ba\label{jan202}
q_{ab}=E\delta_{ab}+{e^c}_c \delta_{ab}-e_{ab}-e_{ba}+O(2) =: E\delta_{ab}+h_{ab}+O(2) \q . 
\ea
Hence $e_{ab}+e_{ba}=-h_{ab}+h_{cd}\delta^{cd}\delta_{ab}+O(2)$ and together with $[[e_{ab}]]=[[e_{ba}]]$ from the Gauss clock (in \ref{choceLonga}) we obtain
\ba\label{jan203}
2\,\,{}^{(1)}[[T^{Da}]] 
&=& {}^{(1)}[[  W^{-2} ( -\tfrac{1}{2} W^{-2}\partial^a\partial_d\partial_e +\tfrac{1}{2} \partial^a \delta_{de} -\partial_e \delta^a_d -\partial_d \delta^a_e)(h^{de}-\delta^{de}h^c_c)]] \nn\\
&=&
 {}^{(1)}[[  W^{-2} ( -\tfrac{1}{2} W^{-2}\partial^a\partial_d\partial_e +\tfrac{1}{2} \partial^a \delta_{de} -\partial_e \delta^a_d -\partial_d \delta^a_e)h^{de}]] \nn\\
&\simeq& 0  \q .
\ea
This is the longitudinal gauge condition for the three metric (see \cite{MFB}).
Furthermore we can calculate the lapse and shift functions introduced in section \ref{lapseandshift}:
\ba\label{jan211}
{}^{(1)}[[\cn^0(k)]] &\simeq& {}^{(1)}[[-\cn\,E^{-1} \frac{2}{3} {e^b}_b(k)]] \nn\\
{}^{(1)}[[\cn^{Da}(k)]] &\simeq&  0\nn\\
{}^{(1)}[[\cn^{Ga}(k)]] &\simeq&    {}^{(1)}[[ \cn \partial^a\, {e^c}_c(k)]]\q\q 
\ea
where we used the definition $\cn={}^{(0)}\cn^0(0)= ({}^{(0)}\! \{T^0(0),C_0(0)\})^{-1} $. In particular we see that the shift function ${}^{(1)}[[\cn^{Da}(k)]]$ vanishes, which is the other gauge condition in the longitudinal gauge (without coupling to spin one (vector) matter fields). 
Hence we can see our first order observables as gauge invariant extensions of longitudinal gauge restricted functions.

These lapse and shift functions can be used to derive in the same way as in section \ref{scalarmodes} the equations of motions for the scalar matter field ${}^{(1)}[[\phi]](\tau)$:  
\ba\label{jan212}
\frac{\bd}{\bd \tau}\,{}^{(1)}[[\phi]](\tau)
&\simeq&
{}^{(1)}[[\cn(\rho-\tfrac{2}{3}E^{-1}\Pi\,e^{b}_b)]](\tau)       \nn\\
\frac{\bd}{\bd \tau}\,{}^{(1)}[[\rho]](\tau)
&\simeq&
{}^{(1)}[[\cn( E^2 \partial^a\partial_a \phi-\tfrac{1}{3}E^2 V'(\Phi){e^a}_a
-E^3 V''(\Phi) \phi +\tfrac{2}{3}E^2V'(\Phi){e^b}_b       )]](\tau) \q . 
\nn\\ 
\ea
These equations lead to the following wave equation\footnote{
For conformal time, that is if one chooses the clock $T^0(0)$ in such a way that 
${}^{(0)}\cn=[\{T^0(0),{}^{0)}C_0(0)\}]^{-1}=E^{-1}$ 
the wave equation (\ref{jan213}) coincides with the wave equation for the (first order gauge invariant) scalar field in \cite{MFB}. 
(The metric scalar mode $\psi(k)$ used in \cite{MFB} can be computed to be $\psi=-\frac{1}{6}E^{-1}{e^c}_c$. Also the (first order) gauge invariant Bardeen potential \cite{bardeen} is given by $\Psi={}^{(1)}[[-\frac{1}{6}E^{-1}{e^c}_c ]]$.)} 
for the scalar mode:
\ba\label{jan213}
\frac{\bd^2}{\bd \tau^2}\,{}^{(1)}[[\phi]](\tau)
&\simeq&
{}^{(0)}[[\cn]](\tau)\,\frac{\bd}{\bd \tau}\,{}^{(0)}[\cn]](\tau)\,\frac{\bd}{\bd \tau}\,{}^{(1)}[[\phi ]](\tau) +\nn\\
&& {}^{(1)}[[\cn^2( E^2\partial^a\partial_a \phi -\tfrac{1}{3}E^2V'(\Phi){e^a}_a -E^3V''(\Phi)\phi               )  ]]  -\nn\\
&& {}^{(0)}[[ \tfrac{2}{3}\cn\Pi  ]](\tau) \,\frac{\bd}{\bd \tau}\,{}^{(1)}[[E^{-1}{e^b}_b  ]](\tau)-{}^{(1)}[[\tfrac{2}{3}\cn E^{-1}{e^b}_b]](\tau)\,\frac{\bd^2}{\bd \tau^2}\,{}^{(0)}[[\Pi]](\tau) \q\q. \nn\\
\ea

Using ${}^{(1)}[[T^{\hat{K}}(\hat k)]]=0+O(2)$ and ${}^{(1)}[[\tilde C_{\hat K}(\hat k)]]\simeq 0+O(2)$ one can replace the metric mode ${}^{(1)}[[{e^b}_b]]$ in (\ref{jan213}) by some combination of the scalar field ${}^{(1)}[[\phi]]$ and its first $\tau$--derivative, however this does not lead to a simple equation. 

Another possibility is to derive a wave equation for the metric mode ${}^{(1)}[[{e^a}_a]]$ in the same way as for the scalar field mode:
\ba\label{jan214}
\frac{\bd}{\bd \tau}\,{}^{(1)}[[{e^a}_a ]] (\tau)
&\simeq&
{}^{(1)}[[-\cn 4 \beta^2 E^2 {a^a}_a ]](\tau)        \nn\\
\frac{\bd}{\bd \tau}\,{}^{(1)}[[{a^a}_a ]](\tau) 
&\simeq&
{}^{(1)}[[\cn( \delta^a\delta_a {e^c}_c  -4 \beta^2 A^2\,{e^c}_c       +8 \beta^2 AE \, {a^c}_c  + 3 \tfrac{\kappa}{\gamma}E^2 V'(\Phi)\phi           )]](\tau)  \nn\\ 
\ea
Hence we obtain the wave equation
\ba\label{jan125}
\frac{\bd^2}{\bd \tau^2}\,{}^{(1)}[[{e^a}_a ]] (\tau)
&\simeq&
{}^{(0)}[[\cn]](\tau)\,\frac{\bd}{\bd \tau}\,{}^{(0)}[\cn]](\tau)\,\frac{\bd}{\bd \tau}\,{}^{(1)}[[{e^a}_a ]](\tau) +\nn\\
&&{}^{(1)}[[ -\cn^2 4 \beta^2 E^2 (\partial^a\partial_a {e^c}_c  - 4 \beta^2 A^2 {e^c}_c  + 3 \tfrac{\kappa}{\gamma} E^2 V'(\Phi)\phi )]](\tau) \q\q
\ea
where we can replace the matter scalar field by
\ba\label{jan216}
{}^{(1)}[[\phi ]](\tau)\simeq 
{}^{(1)}[[\tfrac{\gamma}{\kappa}\Pi^{-1}\tfrac{1}{3}A{e^c}_c ]](\tau) +
{}^{(0)}[[\tfrac{2}{3}(\cn 4 \beta^2 E)^{-1}]](\tau)
\,  \frac{\bd}{\bd \tau}         \,{}^{(1)}[[ {e^a}_a ]](\tau) \q .
\ea
Using the time evolution of the homogeneous variables $A,E,\Phi,\Pi$ and specializing to conformal time ($\cn=E^{-1}$) we therefore have as the  wave equation\footnote{
This wave equation coincides with the wave equation for the Bardeen potential $\Psi={}^{(1)}[[E^{-1}{e^c}_c ]]$ in \cite{MFB}.
} 
for $[[E^{-1}{e^a}_a]]$ 
\ba\label{jan217}
\frac{\bd^2}{\bd \tau^2}\,{}^{(1)}[[E^{-1}{e^a}_a ]] (\tau)
& \simeq &
{}^{(1)}[[4\beta^2 \partial^a\partial_a \, E^{-1}{e^c}_c  ]](\tau)
+
\frac{\bd^2}{\bd \tau^2}\,{}^{(0)}[[E]](\tau) 
\,{}^{(1)}[[E^{-2}{e^c}_c]](\tau)
+ 
\nn\\
&&
{}^{(0)}[[3 E^{-1}]](\tau)\, 
\frac{\bd}{\bd \tau} \,{}^{(0)}[[E]](\tau)\,
\frac{\bd}{\bd \tau} \,{}^{(1)}[[E^{-1}{e^c}_c ]](\tau) -
\nn\\
&&
{}^{(0)}[[2 \Pi^{-1}]](\tau)\, 
\frac{\bd}{\bd \tau} \,{}^{(0)}[[\Pi]](\tau)\,
\frac{\bd}{\bd \tau} \,{}^{(1)}[[E^{-1}{e^c}_c ]](\tau)
-
\nn\\
&&
 {}^{(0)}[[E^{-1}\Pi^{-1}]](\tau)\, 
\frac{\bd}{\bd \tau} \,{}^{(0)}[[E]](\tau)\,
\frac{\bd}{\bd \tau} \,{}^{(0)}[[\Pi]](\tau)\,
\,\,\,{}^{(1)}[[E^{-1}{e^c}_c ]](\tau) \q\q .\q\q\q\q
\ea

\section{First order perturbations: tensor modes} \label{gravitons}

Here we will derive the differential equation for the tensor modes of the gravitational field, that is the $STT$--modes. The $STT$ modes ${}^{STT}\! a^{ab}$ and ${}^{STT}\! e^{ab}$ are already gauge invariants of first order, since they commute with the first order part of the constraints. Hence according to equation (\ref{aaaa}) which connects the first order complete observables with respect to different choices of clock variables, the first order complete observable associated to the $STT$--modes are independent from the choice of clock variables $T^{\hat K}(\hat k)$.

Indeed in the differential equation (\ref{co1}) for the first order complete observables associated to the $STT$ modes (where $\cn^{-1}={}^{(0)}\!\{T^0(0),S(0)\}$)
\ba
\frac{\bd}{\bd \tau} {}^{(1)}[[{}^{STT}a_{ab}]] (\tau) &\simeq&
 {}^{(1)}[[ \{  {}^{STT}a_{ab},\, \tilde C_0(0)\}]](\tau) \,\simeq\,        {}^{(1)}[[ \cn  \{  {}^{STT}a_{ab},\,  S(0)\}]](\tau) \nn\\
\frac{\bd}{\bd \tau} {}^{(1)}[[{}^{STT}e_{ab}]] (\tau) 
&\simeq& {}^{(1)}[[ \{  {}^{STT}e_{ab},\, \tilde C_0(0)\}]](\tau) \,\simeq \,       {}^{(1)}[[ \cn  \{  {}^{STT}e_{ab},\,  S(0)\}]](\tau)
\ea
all terms which may depend on the choice of the $T^{\hat K}(\hat k)$ variables drop out. Here we used the definition (\ref{dec11c}) of $\tilde C_0(0)$ and the fact that the $STT$--modes commute with the first order part of the constraints. 

With the second order scalar constraint (\ref{scalar2nd}) we get
\ba\label{diens1}
\frac{\bd}{\bd \tau} {}^{(1)}\![[{}^{STT}\!a_{ab}]] 
&\simeq&
{}^{(1)}\![[\cn \big( 2\beta E D^{fe}_{ab}\,\,{}^{STT}\! a_{fe}  -2(\beta^2 A^2 
+\tfrac{1}{2}E\tfrac{\kappa}{\gamma}V(\Phi))\,\,{}^{STT}\!e_{ab}+2\beta^2 AE \,\,{}^{STT}\!a_{ab}                \big) ]] \nn\\
\frac{\bd}{\bd \tau} {}^{(1)}\! [[{}^{STT}\! e_{ab}]] 
&\simeq&
{}^{(1)}\! [[ \cn\big(
-2\beta E D^{fe}_{ab}\,\,{}^{STT}\!e_{fe}+2\beta^2E^2\,\,{}^{STT}\! a_{ab}-2\beta^2 AE {}^{STT}\!e_{ab}
\big)]] \q
\ea
where $D^{fe}_{ab}=\frac{1}{2}\epsilon^{cde}\partial_c(\delta_{db}\delta^f_a+\delta_{da}\delta^f_b)$. If $D$ acts on $STT$--modes it simplifies to 
 $D^{fe}_{ab}={\epsilon_b}^{ec}\partial_c \delta^f_a$, moreover on $STT$-modes we have $(D\cdot D)^{fe}_{ab}=-\partial^c\partial_c \, \delta^f_a \delta^e_b$.

We want to derive a wave equation for the ${}^{STT}e_{ab}$ modes. To this end we have to calculate the second ($\tau$--) time derivative of ${}^{STT}e_{ab}$. In the process we have to take also into account the $\tau$ dependence of the homogeneous variables and to replace the ${}^{STT}a_{ab}$--modes by the $\tau$--derivative of the ${}^{STT}e_{ab}$ modes with the help of the second of the equations (\ref{diens1}). The resulting equation of motion for the ${}^{STT}e_{ab}$ modes is
\ba
\frac{\bd^2}{\bd \tau^2} {}^{(1)}\![[{}^{STT}e_{ab}]] 
&\simeq&
\big( {}^{(0)}\! [[
\cn^{-1}]] \frac{\bd}{\bd \tau}{}^{(0)}[[\cn]]-{}^{(0)}[[\cn8\beta^2AE]]\,\big)
\frac{\bd}{\bd \tau} \,{}^{(1)}[[{}^{STT}\! e_{ab}]] \,\,   + \nn\\
&& 
{}^{(1)}\! [[ - \cn^2 (4 \beta^2 E^2 \partial^c\partial_c  
\, +\,
4\beta^2E^3
\tfrac{\kappa}{\gamma}V(\Phi)
+16\beta^4A^2E^2){}^{STT}e_{ab}
]] \q .\q
\ea

The wave equation simplifies if we consider instead of ${}^{STT}\! e_{ab}$ the rescaled variable $E^{-1}\,{}^{STT}\!e_{ab}$: (The correspondence to the metric variables is given by $E^{-1}{}^{TT}\! h_{ab}=-2E^{-1}{}^{STT}\! e_{ab}$ where $h_{ab}$ is the deviation of the spatial metric $q_{ab}$ from the isotropic and homogeneous background $q_{ab}=E\delta_{ab}+h_{ab}=:Q_{ab}+h_{ab}$. Hence the rescaled variable corresponds to $(-2)Q^{ac}\, {}^{TT}\! h_{cb}$.)
%
%
%
%
\ba\label{dienstag2}
\frac{\bd^2}{\bd \tau^2} {}^{(1)}\! [[E^{-1}\,{}^{STT} \! e_{ab}]] &\simeq& 
{}^{(0)}\![[\cn^{-1}]] \frac{\bd}{\bd \tau}{}^{(0)}\! [[\cn]]
 \frac{\bd}{\bd \tau} \,{}^{(1)}\! [[E^{-1}\,{}^{STT} \!\! e_{ab}]]
\,-\,  
{}^{(1)}\! [[
\cn^2 4 \beta^2  E^2 \, \partial^c \partial_c  E^{-1}\,\, {}^{STT}\!\! e_{ab}
]] \, .
 \nn\\
\ea
Specializing to conformal time ($\cn=E^{-1}$) we find 
\ba
\frac{\bd^2}{\bd \tau^2} {}^{(1)}\! [[E^{-1}\,{}^{STT} \! e_{ab}]] &\simeq& 
-{}^{(0)}\![[E^{-1}]] \frac{\bd}{\bd \tau}{}^{(0)}\! [[E]]
 \frac{\bd}{\bd \tau} \,{}^{(1)}\! [[E^{-1}\,{}^{STT} \!\! e_{ab}]]
\,-\,  4 \beta^2   \partial^c \partial_c
{}^{(1)}\! [[
   E^{-1}\,\, {}^{STT}\!\! e_{ab}
]] \,  .  \q 
\ea

\section{Linearization Instabilities}\label{lininstab}

So far we left out the discussion of the integrated diffeomorphism constraints, which start with second order terms (if one does not use three massless scalar fields as clocks for the diffeomorphism constraints). 

Here we will show for a general first class constraint system, 
that one can apply the complete observable method with respect 
to a subset of the constraints, also if these do not form a subalgebra. 
In our case this subset is given by all the constraints with the 
exception of the integrated diffeomorphism constraints. After constructing these (partially invariant) complete observables one can use these to get fully invariant observables.

We will explain the procedure for a finite dimensional system, the generalization to field systems is straightforward. To start consider a first class constraint system with constraints $\{C_j\}_{j=1}^N$ and subdivide this set into two subsets of constraints $\{\{C_a\}_{a=1}^M,\,\{C_\alpha\}_{\alpha=1}^{N-M}\}$. We want to apply the complete observable method to the first subset $\{C_a\}_{a=1}^M$ of the constraint set. Hence we choose a set of clock variables $\{T^A\}_{A=1}^{M}$ and define in the same way as we would deal with the full set of constraints the new constraints
\ba\label{jan261}
\tilde C_A:=(A^{-1})^a_A C_a \q\q \text{where}\q\q \q A^A_a=\{T^A,C_a\}
\ea
These new constraints satisfy
\ba\label{jan262}
\{T^B, \tilde C_A\}=\delta^B_A +\lambda^{BC}_A \tilde C_C
\ea
for some phase space functions $\lambda^{BC}_A$.

Furthermore we want to add to the constraints $\{\tilde C_A\}$ another set of constraints such that we get a complete system of constraints and such that these added constraints commute at least on the constraint hypersurface with the clock variables $\{T^A\}$. Therefore we define the constraints 
\ba\label{jan263}
\tilde C_\alpha &:=& C_\alpha -\{T^A,C_\alpha\}\tilde C_A
\ea
which satisfy
\ba\label{jan264}
\{T^B, \tilde C_\alpha\} &=& O(C)  \q .
\ea

We want to consider the Poisson algebra of the new constraints $\tilde C_A$ and $\tilde C_\alpha$. To this end we define the structure functions $\tilde f$ by
\ba\label{jan265}
\{\tilde C_A,\tilde C_B\} &=& \tilde f_{AB}^C \tilde C_C +\tilde f_{AB}^{\alpha}\tilde C_\alpha  \nn\\
\{\tilde C_A, \tilde C_\beta\} &=& \tilde f_{A\beta}^C \tilde C_C +\tilde f_{A\beta}^{\alpha}\tilde C_\alpha  \nn\\
\{\tilde C_\gamma, \tilde C_\beta\} &=&\tilde f_{\gamma\beta}^C \tilde C_C +\tilde f_{\gamma\beta}^{\alpha}\tilde C_\alpha   \q \q .
\ea
Then using the Jacobi identity we can calculate
\ba\label{jan266}
O(C)=\{\tilde C_A,\{\tilde C_B,T^C\}\} &=&
\{T^C,\{\tilde C_B,\tilde C_A\}\}
+ \{ \tilde C_B,\{\tilde C_A,T^C\}\}  \nn\\
&=&
\{T^C, \tilde f_{BA}^D \tilde C_D +\tilde f_{BA}^\alpha \tilde C_\alpha\}
+
\{\tilde C_B, -\delta^C_A +O(C)\}  \nn\\
&=&
\tilde f_{BA}^C +O(C)  \q .
\ea
Hence we have $\tilde f_{AB}^C=O(C)$ and therefore
\ba\label{jan267}
\{\tilde C_A,\tilde C_B\}=O(C^2)+\tilde f_{AB}^\alpha \tilde C_\alpha \q .
\ea

Similarly 
\ba\label{jan268}
\{T^A,\{\tilde C_B, \tilde C_\alpha\}\}
&=& \{T^A, \tilde f_{B\alpha}^C\tilde C_C +\tilde f_{B\alpha}^\beta \tilde C_\beta \}
\nn\\
&=&
\tilde f_{B\alpha}^A +O(C)  \q .
\ea
On the other hand
\ba\label{jan269}
\{T^A,\{\tilde C_B, \tilde C_\alpha\}\}&=&
\{\tilde C_B,\{T^A, \tilde C_\alpha\}\} + \{\tilde C_\alpha , \{\tilde C_B,T^A\}\} =O(C)
\ea
hence $\tilde f_{B\alpha}^A =O(C)$ and we have that
\ba\label{jan2610}
\{\tilde C_B, \tilde C_\alpha\} &=& O(C^2)+ \tilde f_{B\alpha}^\gamma \tilde C_\gamma \q .
\ea
In the same way one can proof that
\ba\label{jan2611}
\{\tilde C_\alpha, \tilde C_\beta\}&=& O(C^2)+\tilde f_{\alpha\beta}^\gamma \tilde C_\gamma \q .
\ea

In summary we learn that the constraints $\{\tilde C_A\}$ are weakly Abelian modulo terms proportional to the constraints $\tilde C_\alpha$ and that the constraint set $\{\tilde C_\alpha\}$ forms an ideal (modulo terms quadratic in the constraints) of the full constraint algebra.

Assume that we have a phase space function $f$ that is (weakly) invariant under the constraints $\{\tilde C_\alpha\}$. Then we can use the power series for complete observables (\ref{feb035}) just with the constraints $\{\tilde C_A\}$ and compute the complete observable associated to $f$:
\ba\label{jan2612} 
F_{[f;T^A]}(\tau^A) \simeq \sum_{k=0}^\infty
 \frac{1}{k!} 
\{\cdots \{f,\tilde C_{A_1}\}, \cdots \},\tilde C_{A_k}\}
(\tau^{A_1}-T^{A_1})\cdots (\tau^{A_k}-T^{A_k})  \q .
\ea
Because of the properties (\ref{jan267},\ref{jan2610}) the resulting function is (weakly) invariant under both sets of constraints $\{\tilde C_A\}$ and $\{\tilde C_\alpha\}$. (Just consider the Poisson bracket of (\ref{jan2612}) with a constraint $\tilde C_A$ and $\tilde C_\alpha$.) 

That is to find a fully gauge invariant observable we have to start with a partial observable $f$ that is invariant under the constraints $\{\tilde C_\alpha\}$. However it is also possible to calculate complete observables (\ref{jan2612}) associated to non--invariant functions $f$ and then to find fully gauge invariant observables using that
\ba\label{jan2612a}
F_{[f_1\cdot f_2 +f_3;T^A]}(\tau^A) &\simeq &F_{[f_1;T^A]}(\tau^A)\cdot F_{[f_2 ;T^A]}(\tau^A)+F_{[f_3;T^A]}(\tau^A) \q .
\ea
Hence we can obtain a fully gauge invariant observable by combining complete observables associated to non--invariant partial observables $f_i$ algebraically such that the same algebraic combination of the $f_i$ is invariant under the constraints $\tilde C_\alpha$. 

Applied to our situation this means that we can ignore the integrated diffeomorphism constraints in computing complete observables. However in the end we have to find  combinations of partial observables that are invariant under the (altered) integrated diffeomorphism constraints $\tilde C_{Da}(0)$ defined by
\ba\label{jan2613} 
\tilde C_{Da}(0)=C_{Da}(0)-\sum_{ K, k} \{T^{ K}( k), C_{Da}(0)\} \, \tilde C_{ K}( k)  \q\q .
\ea
(The sum does not include the values $(K,k)=(Db,0)$.) Note that also the altered integrated diffeomorphism constraints start at second order.
 
Since the integrated diffeomorphism constraints generated coordinate translations, functions of the form $f(k)g(-k)$ for some $k$ are invariant under the constraints $C_{Da}(0)$. Here we denote by $f(k)$ the Fourier transformation of a field variable $f(\sigma)$. Therefore such functions are a good point to start to look for functions which are invariant under the constraints (\ref{jan2613}).

Concerning the discussion of the gauge invariant $\tau$--generators $H_0(\tau)=F_{[-P_0\, ,T^K]}(\tau)$ in section \ref{clocks}, 
the results do not change if we demand that $f$ in equation (\ref{moon1}) and $P_0$ are invariant (to the order in question) under the altered integrated diffeomorphism constraints $\tilde C_{Da}(0)$.

\end{appendix}

\vspace{1cm}
~\\
{\large{\bf Acknowledgements}}
\vspace{0.2cm}
~\\
We are grateful to Thomas Thiemann for discussions and for suggesting to consider the Bianchi--I model. Furthermore we would like to thank Laurent Freidel and Stefan Hofmann for discussions.  
JT thanks the German National Merit Foundation for financial support and the Perimeter Institute for Theoretical Physics for hospitality. 
Research at Perimeter Institute for Theoretical Physics is supported in part by the Government of Canada through NSERC and by the Province of Ontario through MRI.

\end{document}